\documentclass[namedreferences]{solarphysics}
\pdfoutput=1

\usepackage[hyperref,optionalrh,natbib]{spr-sola-addons} 
\usepackage{graphicx}        
\usepackage{color}           
\usepackage{breakurl}        

\renewcommand{\vec}[1]{{\mathbfit #1}}

\begin{document}

\begin{article}

\begin{opening}

\title{Interpreting the \textit{Helioseismic and Magnetic Imager} 
(HMI) Multi-Height Velocity Measurements}

%
\author{Kaori~\surname{Nagashima}$^{1}$\sep
Bj\"orn~\surname{L\"optien}$^{2}$\sep
   Laurent~\surname{Gizon}$^{1,2}$\sep
Aaron~C.~\surname{Birch}$^{1}$\sep 
Robert~\surname{Cameron}$^{1}$\sep
Sebastien~\surname{Couvidat}$^{3}$\sep
Sanja~\surname{Danilovic}$^{1}$\sep
Bernhard~\surname{Fleck}$^{4}$\sep
Robert~\surname{Stein}$^{5}$}

%
\runningauthor{K.~Nagashima \textit{et al.}}
\runningtitle{HMI Multi-Height Velocity Measurements}

%
  \institute{$^{1}$ Max-Planck-Institut f\"{u}r Sonnensystemforschung, 
Justus-von-Liebig-Weg 3, 37077 G\"ottingen, Germany \\
                     email: \url{nagashima@mps.mpg.de} \\ 
             $^{2}$ Institut f\"ur Astrophysik, Georg-August-Universit\"at G\"ottingen, 
37077 G\"ottingen, Germany \\
 $^{3}$ Stanford University, Stanford, CA 94305, USA \\
 $^{4}$ ESA Science Operations Department, c/o NASA/GSFC, Greenbelt, MD 20771, USA \\
 $^{5}$ Michigan State University, East Lansing, MI 48824, USA
             }

\begin{abstract}
The \textit{Solar Dynamics Observatory/Helioseismic and Magnetic Imager} 
(SDO/HMI) filtergrams, taken at six wavelengths around 
the Fe \textsc{i} 6173.3 \AA \ line, contain information about the line-of-sight
velocity over a range of heights in the solar atmosphere. 
Multi-height velocity inferences from these observations can 
be exploited to study wave motions and energy transport in the atmosphere.
Using realistic convection simulation datasets provided by the
\textsf{STAGGER} and \textsf{MURaM} codes, 
we generate synthetic filtergrams and
explore several methods for estimating Dopplergrams.
We investigate at which height each synthetic Dopplergram
correlates most strongly with the vertical velocity in the model atmospheres.
On the basis of the investigation, we propose two Dopplergrams
other than the standard HMI-algorithm Dopplergram produced from HMI filtergrams:
a line-center Dopplergram and an average-wing Dopplergram.
These two Dopplergrams correlate most strongly with vertical
velocities at the heights of 30\,--\,40 km above (line-center)
and 30\,--\,40 km below (average-wing) 
the effective height of the HMI-algorithm 
Dopplergram. Therefore, we can obtain velocity information from 
two layers separated by about a half of a scale height in the atmosphere, 
at best. 
The phase shifts between these multi-height Dopplergrams from observational data
as well as those from the simulated data are also
consistent with the height-difference estimates 
in the frequency range above the photospheric acoustic cutoff frequency. 
\end{abstract}

%

\end{opening}

%

\section{Introduction}
In recent helioseismology studies
multi-height velocity and intensity information have been used 
in addition to standard photospheric Dopplergrams.
\inlinecite{2008A&A...481L...1M} investigated the phase shift 
between photospheric- and chromospheric-intensity datasets obtained 
by the \textit{Hinode} satellite. They reported large 
phase differences along the $p$-mode ridges and no phase difference
on the $f$-mode ridge.
\inlinecite{2009ApJ...694L.115N} inferred chromospheric downflows
by interpreting multi-height observations.
More recently, 
\inlinecite{2012SoPh..281..533H} examined the phase differences between
several observables originating from various layers obtained by the
\textit{Helioseismic and Magnetic Imager} (HMI: \opencite{2012SoPh..275..207S}) 
and \textit{the Atmospheric Imaging Assembly} 
(AIA: \opencite{2012SoPh..275...17L}) 
onboard the \textit{Solar Dynamics Observatory} (SDO: \opencite{2012SoPh..275....3P}).
\inlinecite{2012SoPh..tmp..303R} exploited multi-height HMI and AIA data 
to study power enhancement around active regions at various heights in the atmosphere.
They used not only the standard HMI-algorithm Dopplergrams but 
also Doppler information derived from the line wing 
(this is similar to what we define as the ``far-wing" Dopplergram in 
Section \ref{sec:DopDef}). 

Multi-height information is, however, useful not only 
for helioseismology studies but also for many other research purposes,
for example, the study of energy transport in the solar atmosphere 
(\textit{e.g.} \opencite{2006ApJ...648L.151J}, \opencite{2008ApJ...681L.125S},
\opencite{2009ASPC..415...95S}, \opencite{2010ApJ...723L.134B},
\opencite{2011A&A...532A.111K}).
If we could obtain multi-height velocity information from SDO/HMI datasets,
we would have huge datasets available compared with
other current instruments; HMI obtains full-disk datasets without
significant interruptions.

Here we show that we can obtain multi-height velocity information 
from SDO/HMI observations;
we use realistic numerical convection simulations to characterize these 
multi-height Dopplergrams.
In Section \ref{sec:makeDop} we introduce HMI observables as well as the 
simulation datasets,
and define several types of Dopplergrams. 
In Section \ref{sec:syn_cor}
we investigate the contribution heights of the Doppler velocities 
using realistic convection simulations. 
On the basis of these contribution height estimates and 
availability of the observables we choose 
a pair of Doppler velocities as rather robust multi-height velocity datasets;
this is summarized in Section \ref{sec:multi-height}.
We measure the phase difference
derived from the HMI observation datasets as well as
simulated datasets in Section \ref{sec:phase}.
Finally, a brief summary is given in Section \ref{sec:Conclusions}.

\section{Making Multi-Height Dopplergrams from SDO/HMI Datasets} \label{sec:makeDop}

\subsection{HMI Observables} \label{sec:HMI}

\begin{figure}
\centerline{\includegraphics[]{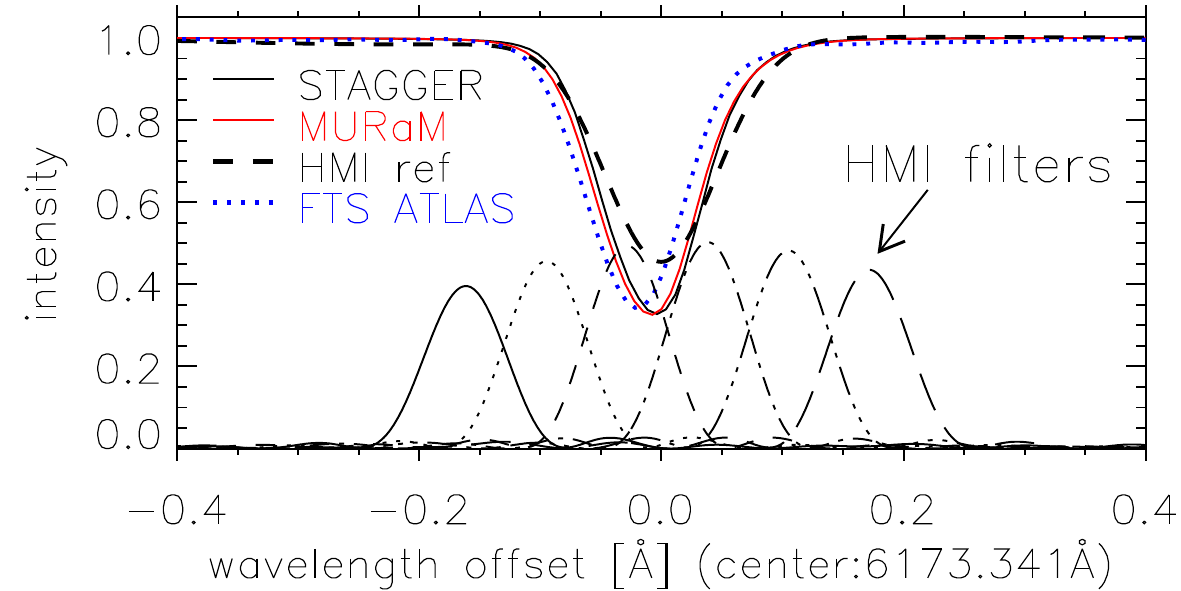}}
\caption{Profile of the Fe \textsc{i} absorption line at 6173.3 \AA. 
The HMI reference line profile, NSO FTS atlas profile, 
synthetic line profiles of \textsf{STAGGER} and \textsf{MURaM} data averaged over each FOV, and
HMI filter profiles are shown. The central wavelength positions of the filter are 
$+172$ m\AA ($\lambda_0$, long-dashed), 
$+103.2$ m\AA ($\lambda_1$, dash-triple-dotted), 
$+34.4$ m\mbox{\AA} ($\lambda_2$, dash-dotted), 
$-34.4$ m\mbox{\AA} ($\lambda_3$, short-dashed), 
$-103.2$ m\mbox{\AA} ($\lambda_4$, dotted), 
and $-172$ m\mbox{\AA} ($\lambda_5$, solid)
from right to left.}\label{fig:prof_line}
\end{figure}

SDO/HMI takes full-disk images in the Fe \textsc{i} absorption 
line at $\lambda_{\mathrm lc}= 6173.3433$ \AA \  (at rest). 
Every 45 seconds 12 filtergrams are taken at left- and right-circular polarizations 
at six wavelength positions around the line center:
$+172$ m\mbox{\AA} ($\lambda_0$), 
$+103.2$ m\mbox{\AA} ($\lambda_1$), $+34.4$ m\mbox{\AA}  ($\lambda_2$), 
$-34.4$ m\mbox{\AA}  ($\lambda_3$), $-103.2$ m\mbox{\AA}  ($\lambda_4$), 
and $-172$ m\mbox{\AA} ($\lambda_5$).
Example filter profiles and the reference line profile 
used in the standard HMI data processing are shown 
in Figure \ref{fig:prof_line} (see \opencite{2012SoPh..278..217C} for details).
The HMI reference line profile is a simple combination of 
Voigt and Gaussian functions which are calibrated for the purpose of 
obtaining Dopplergrams (\opencite{SDOlinePrep}).
Figure \ref{fig:prof_line} also shows the National Solar Observatory (NSO) 
\textit{Fourier Transform Spectrometer} (FTS) Atlas spectra 
(\opencite{1998assp.book.....W}),
the reference line profile used in the HMI data processing 
and also the synthesized line profiles (Section \ref{sec:syn_sp_filt}).
The pixel size is 0.505 arcsec, or 370 km on the Sun at the disk center.
In this article we consider only the data obtained by the front camera, 
which is the camera for the line-of-sight observables.

The standard Dopplergram products of the HMI pipeline 
(hereafter, HMI-algorithm Dopplergrams)
are made from these sets of filtergrams
(\opencite{2012SoPh..278..217C}),
and the formation height of the HMI-algorithm Dopplergram is around 100 km 
above $\tau_{5000 \mbox{\AA}} =1$ 
(\opencite{2011SoPh..271...27F}),
while the line-core formation height is estimated to be around 200--300 km 
(\textit{e.g.} \opencite{2006SoPh..239...69N}).

In this study, we show that 
by combining the HMI filtergrams at six wavelengths in several ways 
we can derive multi-height velocity information in the solar atmosphere.
Note that we always take the sum of the intensities at left- and right-circular polarizations 
at every wavelength, and hereafter we use only these total intensities 
[$I_i$ ($i=0,1, \dots, 5$) at $\lambda_i$].
In the HMI pipeline, on the other hand, HMI-algorithm 
Doppler velocities are computed separately from the left- and right-circular polarization intensities, and then the two Doppler velocities are averaged.
Since the data processing is non-linear, this might cause
small differences from our approach. In this study, however, we limit ourselves to 
quiet-Sun data. Thus the two approaches should produce 
similar results.

\subsection{Realistic 3D Convection Simulation Datasets}
\begin{figure} [htp]
\centerline{\includegraphics[]{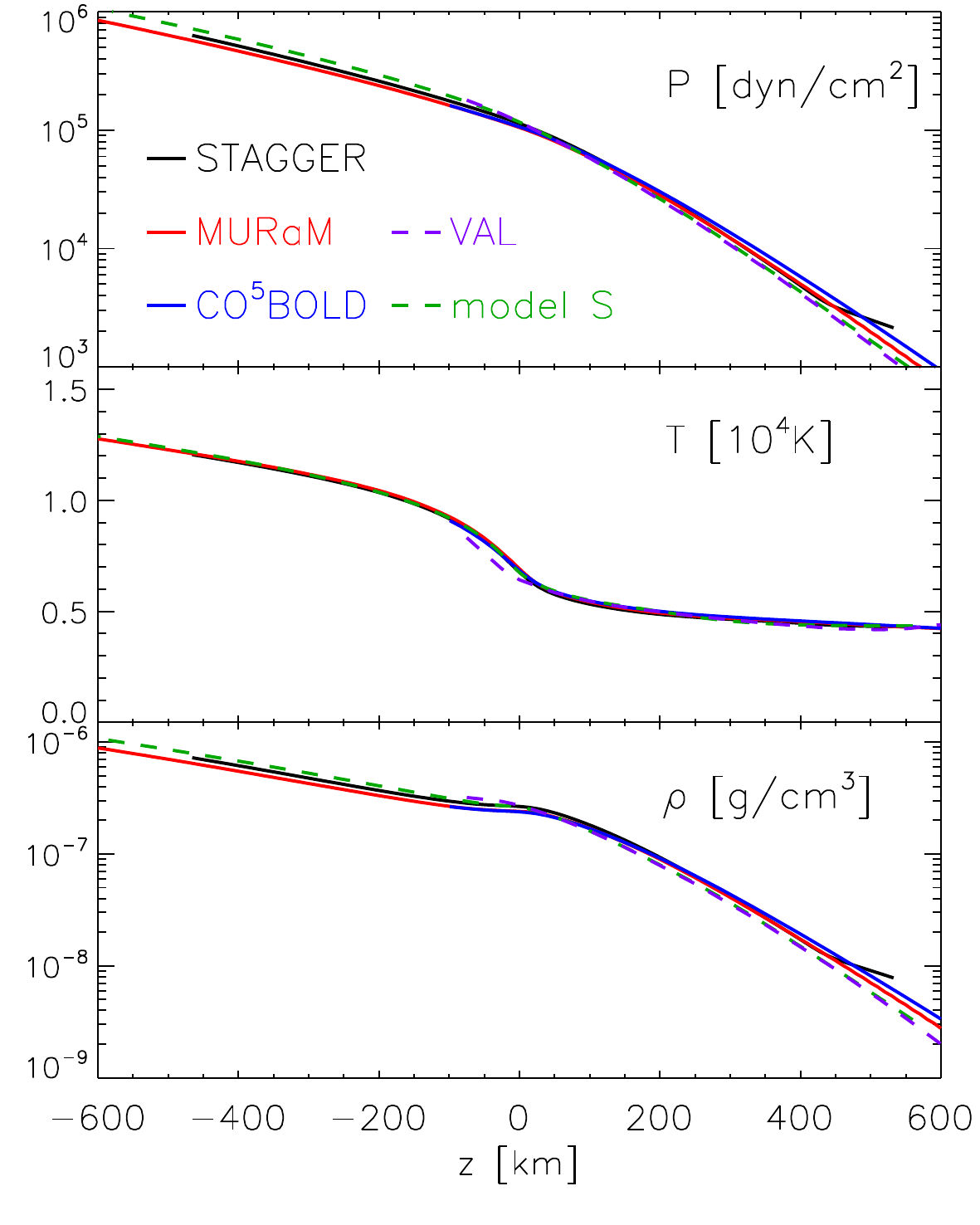}}
\caption{Average atmospheric profiles of \textsf{STAGGER} (black solid),  
\textsf{MURaM} (red solid), and \textsf{CO$^5$BOLD} (blue solid) 
simulations as well as Model S (green dashed) and VAL (purple dashed). 
The zero point of the 
height $z$ is defined as the height where $\tau_{5000\mbox{\AA}}=1$ .
}\label{fig:prof_atmo}
\end{figure}

To estimate the contribution heights of ``multi-height" Dopplergrams, 
we use datasets from two different simulation codes: 
\textsf{STAGGER} (\opencite{2009AIPC.1094..764S}; 
\citeyear{2009ASPC..416..421S}; \opencite{2012LRSP....9....4S}) and 
\textsf{Max-Planck-Institut f\"{u}r Sonnensystemforschung/University 
of Chicago RAdiative MHD}
(\textsf{MURaM}: \opencite{2005A&A...429..335V}) code. 
The average profiles of the solar atmosphere models in these simulations
are shown in Figure \ref{fig:prof_atmo} as well as 
a standard solar model (Model S: \opencite{1996Sci...272.1286C})
and the VAL atmosphere (\opencite{1981ApJS...45..635V}) for comparison.
Since we also compare our results with the results from 
\textsf{COnservative COde for the COmputation of COmpressible COnvection in a BOx of L Dimensions with l=2,3} 
(\textsf{CO$^5$BOLD}: \opencite{2002AN....323..213F};
\opencite{2004A&A...414.1121W})
at several points, we show these profiles as well. 
The height [$z$] is defined relative to the layer where $\tau_{5000\mbox{\AA}}=1$.
The differences in the mean structure in the three simulations 
relative to Model S are within 10\,\% at the surface.
However, at 400 km above the surface
the pressure and the density differences between the models 
are up to 30\,\% and 25\,\% while the temperature difference is 5\,\%.
A detailed comparison of these models is given by
\cite{2012A&A...539A.121B}.

\subsubsection{Simulated Data}
\paragraph{\textsf{STAGGER} Data}

We use a 3D convection simulation provided by the \textsf{STAGGER} code. 
The simulation domain is 96 Mm wide and extends vertically 
from the temperature minimum down to a 
depth of 20 Mm. There is only weak photospheric magnetic field;  
the average unsigned vertical field is 3 G, the average unsigned horizontal field is 5 G,
and the maximum field is 2.2 kG at the surface. 
The horizontal resolution is 48 km, and the vertical
resolution varies from 12 km near the surface to 80 km at large depths.  
From this domain we use only a subregion for the radiative-transfer calculations. 
The subregion is 48 Mm $\times$48 Mm wide and extends from the 550 km above to 450 km 
below the surface with 
$\tau_{5000 \mbox{\AA}}=1$. This $\tau_{5000 \mbox{\AA}}$ is defined by the average atmospheric profiles shown in Figure \ref{fig:prof_atmo}.

\paragraph{\textsf{MURaM} Data}

We also use convection-simulation results provided by 
the \textsf{MURaM} code 
with non-gray radiative transport using four opacity bins 
(\opencite{2004A&A...421..741V}).
The region we use is 9 Mm wide in the horizontal directions and covers
3 Mm in the vertical direction (from 1.98 Mm below the surface up to 1 Mm above the 
$\tau_{5000 \mbox{\AA}} = 1$ level)
with 17.6 km horizontal resolution and 10 km vertical resolution. 
This simulation is non-magnetic.

\subsubsection{Spatial Resolution} \label{sec:sim_resolution}
The original pixel sizes of \textsf{STAGGER} and \textsf{MURaM} simulation datacubes are 
48 km and 18 km, respectively, and are much smaller than the HMI pixel size 
(370 km on the Sun at the disk center), or the HMI diffraction limit (1.83 HMI pixels).
As discussed by \inlinecite{2011SoPh..271...27F}, 
the spatial resolution may affect the analyses.
Here we use a theoretical point spread function (PSF) to synthesize HMI observables 
from the simulation datasets. 
The theoretical PSF [$P$] for an optical system with aperture diameter [$D$] 
and focal length [$f$] at wavelength [$\lambda$] is given by 
\begin{eqnarray}
P(r^{\prime}) = \left[\frac{2J_1 (r^{\prime})}{r^{\prime}} \right]^2, \label{eq:PSF}
\end{eqnarray}
where $J_1$ is the first-order Bessel function of the first kind, 
$r^{\prime}=\pi D r /(\lambda f)$ is 
the normalized radius,
and $r$ is the radial distance from the optical axis on the focal plane 
(\textit{e.g.} \opencite{2002tsai.book.....S}).
The first zero of $P$ is at $r'=1.22\pi$, and the diffraction limit for HMI 
is defined here as
$ r_{\mathrm{dif. limit}}= \lambda f/D = 21.9 \ \mu \mathrm{m} $, where $\lambda=\lambda_{\mathrm{lc}}$ and
the effective focal ratio $f/D=35.42$ (\opencite{2012SoPh..275..229S}).
The HMI CCD pixel size is 12 $\mu$m, so the diffraction limit 
corresponds to 1.83 HMI pixels.

Because of the limited field of view of the \textsf{MURaM} data, 
we apply this PSF to the \textsf{STAGGER} data only. 
If we apply the PSF to the \textsf{MURaM} data, we have  
only $\approx 10^2$ data points in the field of view.

\subsubsection{Synthetic Spectra and Filtergrams} \label{sec:syn_sp_filt}
\begin{figure}[hbt]
\centerline{\includegraphics[]{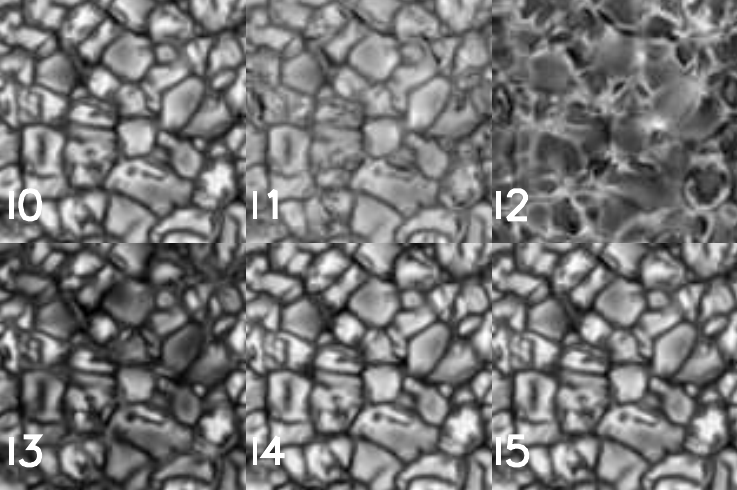}}
     \vspace{0.03\textwidth}  
\centerline{\includegraphics[]{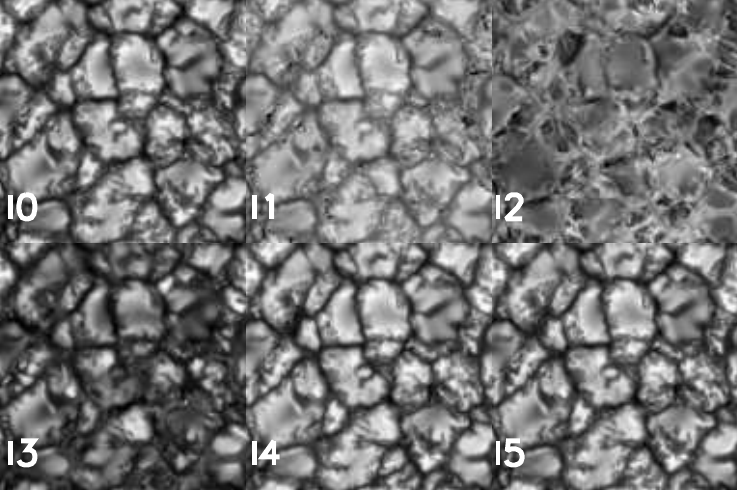}}
\caption{Sample synthetic filtergrams
obtained from the \textsf{STAGGER} datasets (top) and 
\textsf{MURaM} datasets (bottom).
The labels $I_i$ ($i=0,1, \dots, 5$) indicate filtergrams with the filter centered 
at $\lambda_i$.
The area of the \textsf{STAGGER} maps is 10 Mm square, which is about one-fifth of the original 
field of view. The area of the MURaM maps is 9 Mm square, the full original field of view. 
With these original resolutions, the granular pattern is clearly seen in any wavelength, 
although the contrast depends on the wavelength. }
\label{fig:synth_samplefilt}
\end{figure}

\begin{figure}[hbt]
\centerline{\includegraphics[]{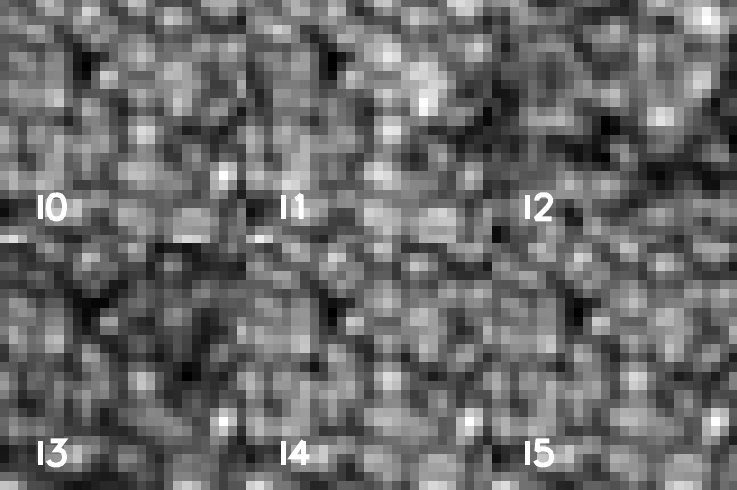}}
     \vspace{0.03\textwidth}  
\centerline{\includegraphics[]{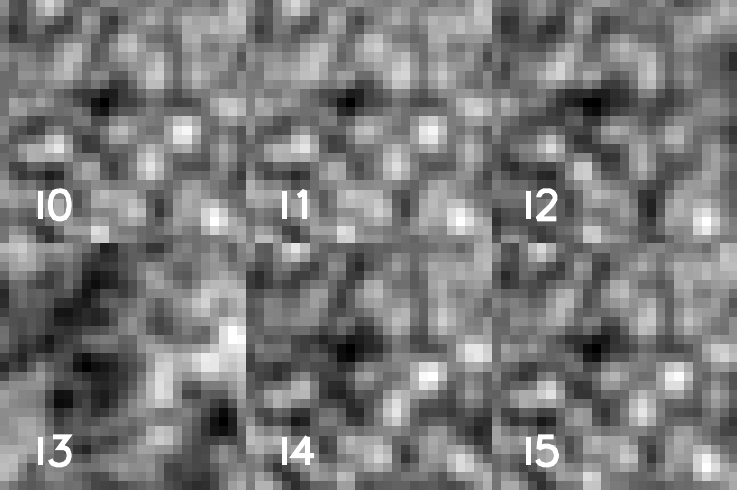}}
\caption{Sample filtergrams obtained from \textsf{STAGGER} datasets with the resolution reduced by an approximate HMI PSF (top) and example HMI observations (bottom).
The labels $I_i$ ($i=0,1, \dots, 5$) indicate filtergrams 
with the filter centered at $\lambda_i$.
The area is 10 Mm square and the field of view of the top panel is identical to
the \textsf{STAGGER} maps (top panels) of Figure \ref{fig:synth_samplefilt}.
The top and bottom panels look reasonably similar. With this resolution, each granular cell structure is marginally 
resolved.}
\label{fig:synthPSF_obs_filtgram}
\end{figure}

Using the simulated datasets,
we calculate the line profile for the Fe \textsc{i} absorption line 
at 6173.3 \AA\ using the 
\textsf{Stokes-Profiles-INversion-O-Routines} (\textsf{SPINOR}) code (\opencite{2000A&A...358.1109F}).
The atomic parameters that we use to synthesize the line profiles are 
summarized in Table \ref{tab:SPINORparam}. The abundance is from 
\inlinecite{2002A&A...391..331B}.
Since this line is formed relatively deep in the photosphere, 
the assumption of LTE for the line synthesis calculations seems 
justified for the particular objectives of this article
(\textit{cf.}, \textit{e.g.}, \opencite{2009A&A...494.1091B}; \opencite{2011SoPh..271...27F}).
The average line profiles obtained from \textsf{STAGGER} and \textsf{MURaM} data 
are shown in Figure \ref{fig:prof_line}.
Figure \ref{fig:synth_samplefilt} shows a set of snapshots of synthetic filtergrams 
created from synthetic spectra convolved with the HMI filter profiles, and
Figure \ref{fig:synthPSF_obs_filtgram} shows the filtergrams 
obtained from the \textsf{STAGGER} datasets after the PSF was applied. 
For comparison, Figure \ref{fig:synthPSF_obs_filtgram}
also shows the HMI filtergrams 
taken at 07:29:15 UT on 23 January 2011 near the disk center. 
At the time of the observation, the SDO 
line-of-sight motion toward the Sun was
288.25 m s$^{-1}$
and this is subtracted when we calculate the Doppler velocities from these filtergrams
(shown in Figure \ref{fig:synthPSF_obs_Dop} in Section \ref{sec:DopDef}). 
The filtergrams created from the simulation datasets (upper panels of 
Figure \ref{fig:synthPSF_obs_filtgram}) look
qualitatively similar to the observations 
(lower panels of Figure \ref{fig:synthPSF_obs_filtgram}),
although the fact that the structure sizes in observation datasets 
seem slightly bigger than those in the synthesized datasets
may indicate that the real PSF for HMI is slightly worse than 
the purely theoretical one we used here.

\begin{table}[ht]
 \caption{Atomic parameters for the line synthesis.}\label{tab:SPINORparam}
\begin{tabular}{cccccc}     
\hline
&&& &\multicolumn{2}{c}{Ionization potential} \\
Wavelength & Atomic mass & Fe abundance & log gf & 1st & 2nd \\
\hline
6173.341 \AA & 55.8500 & 7.43 & -2.880 & 2.222 eV & 7.9024 eV \\
\hline
\end{tabular}
\end{table}

\subsection{Several Types of Doppler Velocity Measurements}\label{sec:DopDef}

We calculate several types of velocities from the synthetic line spectra and 
the filtergrams.
Although only six filtergrams are available from the HMI observations,
here we derive these quantities from full spectra as well (5 and 6 in the list below)
for the synthetic data analyses. 
We compute the intensities at 200 points in wavelength 
from $-0.75$ \AA \ to $+0.75$ \AA \ around the line center using \textsf{SPINOR}.
Although there are still some tiny sidelobes in the HMI filters 
outside this range (see \opencite{2012SoPh..278..217C} for details), 
they do not matter for the analyses performed here.
In the calculation of the absorption-line profile, we do not include the 
turbulent velocity in the simulated atmosphere. 
Therefore, the line profile for each pixel shows a narrower and steeper profile
compared with the line profile averaged over the entire FOV (shown in Figure \ref{fig:prof_line}).

\subsubsection{Doppler Velocity 1: Core, Wing, Far-Wing, and Average-Wing Doppler Velocities}
These Doppler velocities are derived from the Doppler signal made 
of pairs of filtergrams, 
\[ D_\mathrm{br}= (I_\mathrm{b}-I_\mathrm{r})/ (I_\mathrm{b}+I_\mathrm{r}) \ ,\] 
where $I_\mathrm{b}$ and $I_\mathrm{r}$ are 
the intensities of the blue and red sides of each pair.
We define the core pair as $(I_\mathrm{r}, I_\mathrm{b})=(I_2, I_3)$, 
the wing pair as $(I_\mathrm{r}, I_\mathrm{b})=(I_1, I_4)$,
the far-wing pair as $(I_\mathrm{r}, I_\mathrm{b})=(I_0,I_5)$,
and the average-wing pair as 
$(I_\mathrm{r},I_\mathrm{b})=([I_0+I_1]/2,[I_4+I_5]/2)$. 
Conversion of the Doppler signal into velocities is discussed 
in Appendix \ref{sec:sim_DopsigFit}.

\subsubsection{Doppler Velocity 2: Center of Gravity (Six Points)}
The Doppler velocity of the center of gravity (cog) of the line is derived from \begin{eqnarray}
v_{\mathrm{cog}}=c \left( \frac{ \sum_{i=0}^{5} \lambda_i \{I_i-I_\mathrm{max}\} }{ \lambda_{\mathrm lc} \sum_{i=0}^{5} \{I_i-I_\mathrm{max} \}} -1 \right),
\end{eqnarray}
where $\lambda_{\mathrm lc}$ is the line-center wavelength, 
$c$ is the speed of light, $\lambda_i$ is the wavelength where the intensity $I_i$
is measured, and $I_\mathrm{max}$ is the maximum of $I_i (i=0, 1, \dots , 5)$.

\subsubsection{Dopler Velocity 3: Line Center (Three Points)}
The Doppler velocity of the line center can be estimated 
from the three wavelength points around the 
minimum-intensity wavelength. 
We calculate the parabola through the three points 
and use the apex of the parabola as an estimate of the line-center shift.

\subsubsection{Doppler Velocity 4: Simplified HMI-Algorithm Dopplergrams}
We calculate simplified HMI-algorithm Doppler velocities
[$v_{\mathrm{HMI1}}$ and  $v_{\mathrm{HMI2}}$]
based on the method used for HMI pipeline products 
(\opencite{2012SoPh..278..217C}).
The procedure here is identical to that of \inlinecite{2011SoPh..271...27F}. 
First we calculate the phase of the first and second 
Fourier coefficients [$\phi_\mathrm{F1}$ and $\phi_\mathrm{F2}$]
of the line profile (made of six filtergrams, namely, 
$I_i$ for $i=0,1, \dots, 5$).
Then we obtain the velocities as $v_{\mathrm{HMI1}} = \alpha \phi_\mathrm{F1}/(2 \pi)$ and
$v_{\mathrm{HMI2}} = \alpha \phi_\mathrm{F2}/(4 \pi)$,
where $\alpha=c/\lambda_{\mathrm lc} (68.8 \mathrm{[m\mbox{\AA}]} \times 6)$
(see \opencite{2012SoPh..278..217C} for details).
Note that the widely-used HMI-algorithm Dopplergram provided by the 
HMI pipeline is the first one [$v_{\mathrm{HMI1}}$], and 
the HMI-algorithm Dopplergrams mentioned in this article are $v_{\mathrm{HMI1}}$
unless otherwise noted.

\subsubsection{Doppler Velocity 5: Center of Gravity (Full)}
The Doppler velocity of the center of gravity of the full line profiles is calculated by
\begin{eqnarray}
v_{\mathrm{cog, f}}=c \left(\frac{ \int_{\lambda_\mathrm{m}}^{\lambda_\mathrm{M}} 
\mathrm{d}\lambda \   \lambda ( I(\lambda) -I_\mathrm{c}) }
{ \lambda_\mathrm{lc} \int_{\lambda_\mathrm{m}}^{\lambda_\mathrm{M}} \mathrm{d}\lambda   ( I(\lambda) -I_\mathrm{c})} -1 \right),
\end{eqnarray}
where $\lambda_\mathrm{m}$ and $\lambda_\mathrm{M}$ are minimum and maximum wavelengths of the profile,
and $I_\mathrm{c}$ is the continuum intensity.
Here $\lambda_\mathrm{m}$ and $\lambda_\mathrm{M}$ are $-0.75$ \AA\ and $+0.75$ \AA\ from the line center, respectively.

\subsubsection{Doppler Velocity 6: Line Center (Full)}
The Doppler velocity of the line center of the full line profile is determined by the location of 
the minimum of a 
fourth-order polynomial fit over the range $\pm 32 \ \mbox{m\AA}$ around the minimum-intensity wavelength.

\subsubsection{Doppler Velocity Maps}
\begin{figure}[htbp]
\centerline{\includegraphics[]{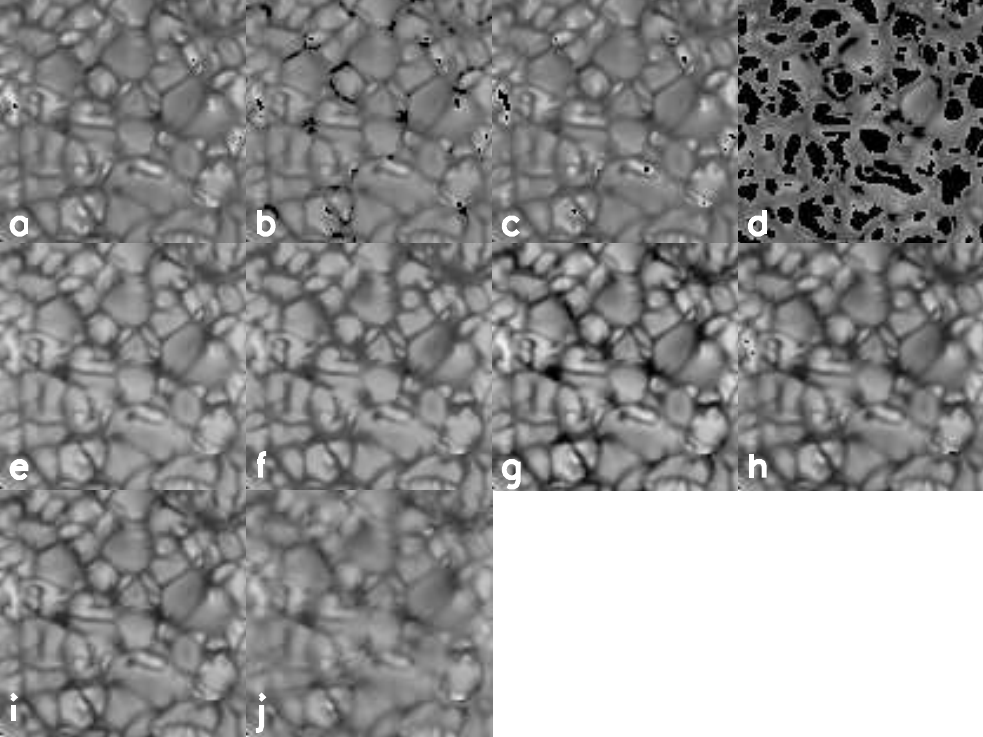}}
 \vspace{0.03\textwidth}  
\centerline{\includegraphics[]{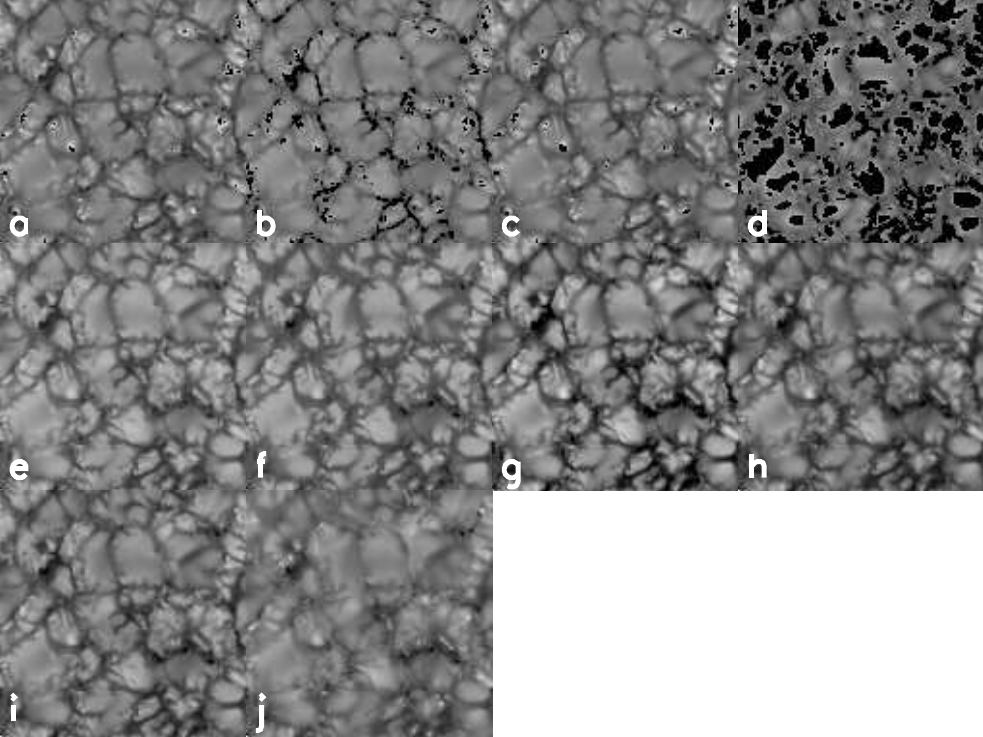}}
\caption{Sample synthetic Dopplergrams
obtained from the \textsf{STAGGER} simulations (top) and from the \textsf{MURaM} simulations (bottom):
average-wing (a), far-wing (b), wing (c), core (d), center of gravity from filtergrams (e),
line center from filtergrams (f), 
first simplified HMI-algorithm Dopplergram (g), 
second simplified HMI-algorithm Dopplergram (h),
center of gravity from full spectra (i), and
line center from full spectra (j).
The areas and the field of view are identical to those of Figure \ref{fig:synth_samplefilt} (top: 10 Mm square, bottom: 9 Mm square).
The gray scale range is from $-5$ (black, downflow) to $+5$ km s$^{-1}$ (white, upflow).
Note that in the panels (b)\,--\,(d), especially in (d), there are many unusable points
(in black) because of the limited applicable range (see Appendix \ref{sec:sim_DopsigFit} for details).}
\label{fig:synth_sampleDop_STMR}
\end{figure}

\begin{figure}[htbp]
\centerline{\includegraphics[]{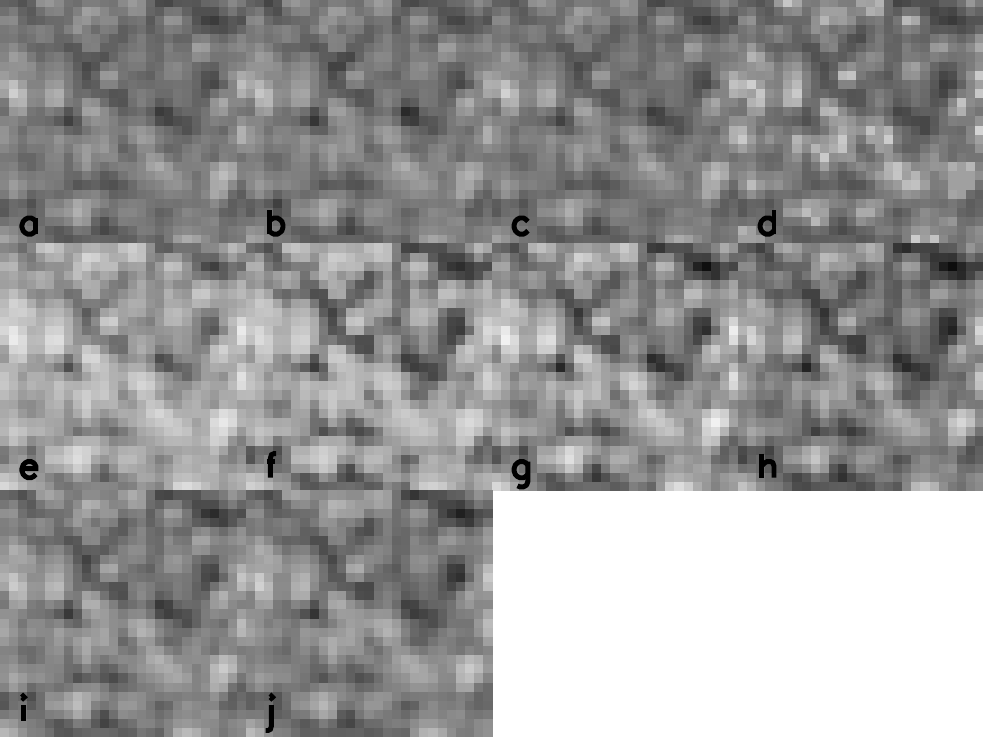}}
\vspace{0.03\textwidth}
\centerline{\includegraphics[]{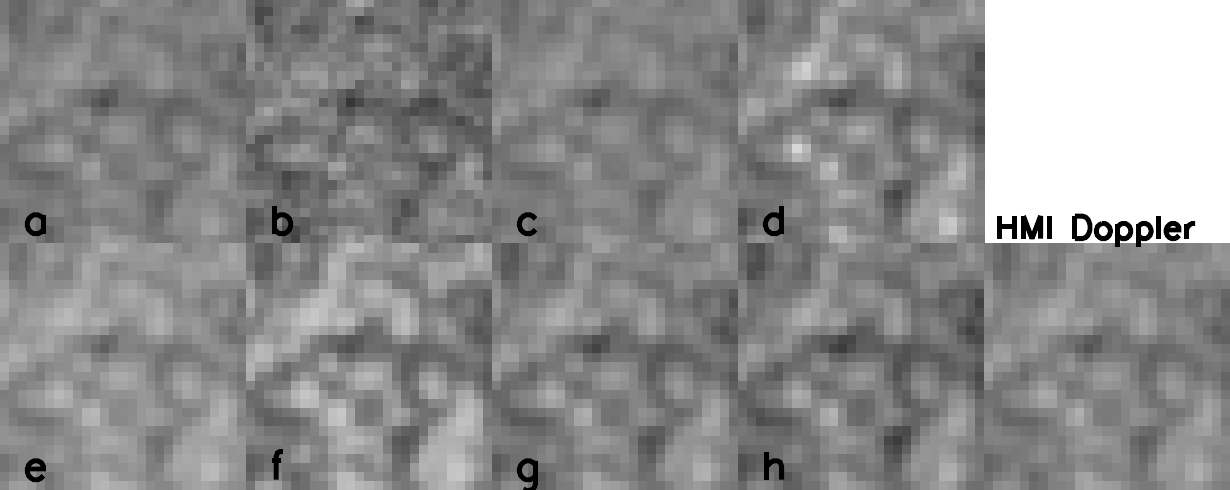}}
\caption{
Sample synthetic Dopplergrams obtained from \textsf{STAGGER} datasets with
the resolution reduced by the approximate HMI PSF (top) and 
obtained from HMI observations (bottom):
average-wing (a), far-wing (b), wing (c), core (d), center of gravity from filtergrams (e),
line center from filtergrams (f),
first simplified HMI-algorithm Dopplergram (g),
second simplified HMI-algorithm Dopplergram (h),
center of gravity from full spectra (i), and
line center from full spectra (j).
The area of each map is 10 Mm squre and the fields of view are identical to 
Figure \ref{fig:synthPSF_obs_filtgram}.
The gray-scale range in this figure 
is from $-2.5$ (black, downflow) to $+2.5$ km s$^{-1}$ (white, upflow), which
is smaller than that in Figure \ref{fig:synth_sampleDop_STMR}.
The contrast of the Doppler velocity is 
significantly reduced by the PSF smoothing.
The HMI-algorithm Dopplergram is also shown in the bottom right panel.}
\label{fig:synthPSF_obs_Dop}
\end{figure}

Samples of these velocity maps are shown in Figure 
\ref{fig:synth_sampleDop_STMR} (raw \textsf{STAGGER} and \textsf{MURaM})
and Figure \ref{fig:synthPSF_obs_Dop} (\textsf{STAGGER} 
with the resolution reduced by the HMI PSF and HMI observation).
For the HMI observation datasets, we lack the center-of-gravity and 
the line-center Doppler velocities from 
the full spectra because we only have filtergrams. Instead, 
we show the standard HMI-algorithm Dopplergram provided by the pipeline as well.
The contrasts in the different Dopplergram observables are different.
This suggests that we might have information from various heights.
It is clear from the comparison between 
Figures \ref{fig:synth_sampleDop_STMR} and \ref{fig:synthPSF_obs_Dop} that
the contrast of the Doppler velocity is significantly reduced 
after smoothing with PSF, although the differences in contrast
between the different Dopplergram observables are still present.
In Figure \ref{fig:synthPSF_obs_Dop} 
the gray-scale contrast, or the dynamic range, of the observation 
Dopplergrams is less than the synthesized one. 
This may also indicate the difference between the theoretical PSF that 
we used and the real PSF for HMI instruments.

\section{Comparison of the Synthetic Doppler Velocities and the Original Velocity Field in the Simulation Box} \label{sec:syn_cor}

\begin{figure}
\centerline{\includegraphics[]{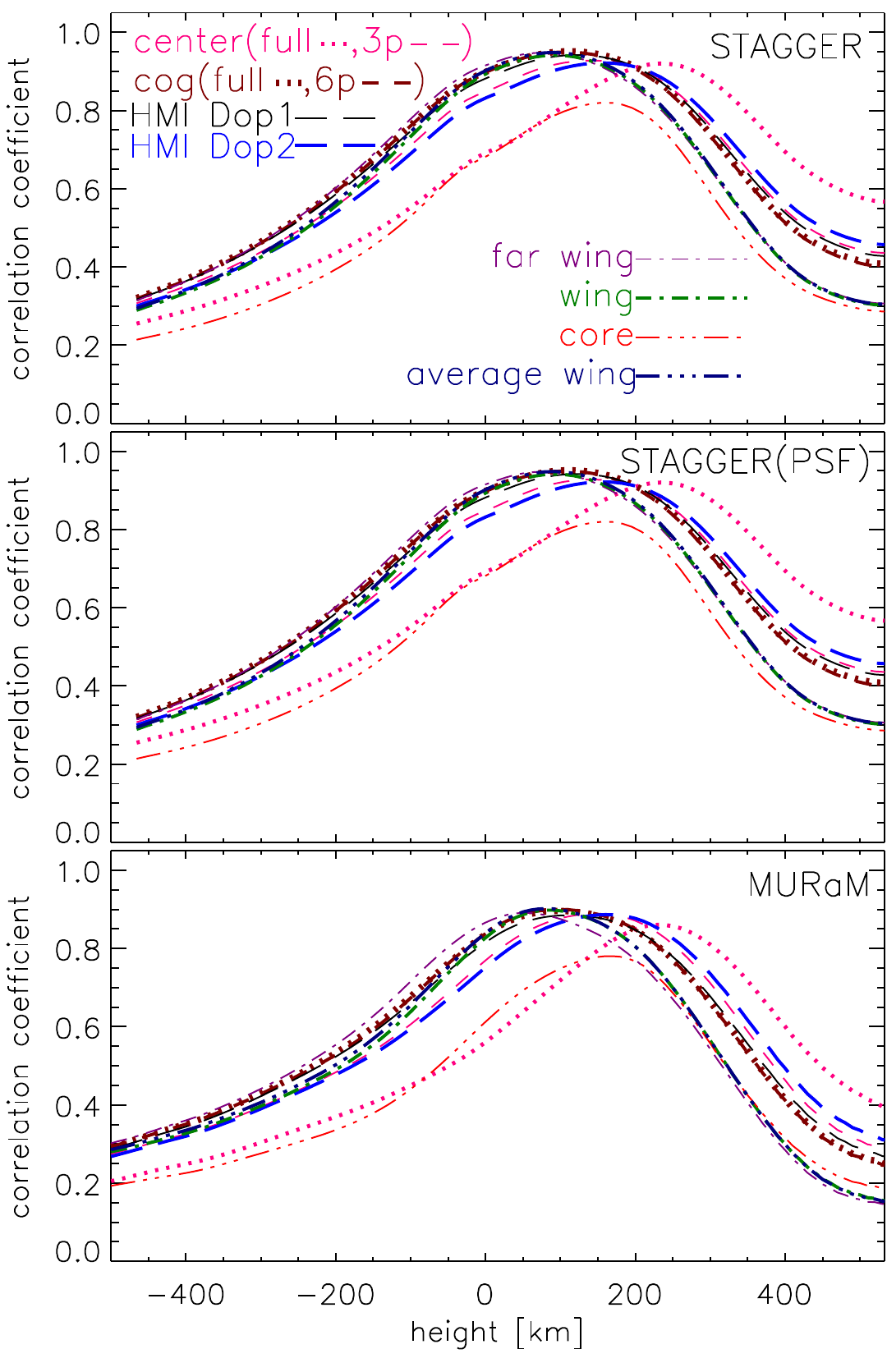}}
\caption{Correlation coefficients of 
the synthetic Doppler velocities and the original $v_z$.
The top, middle, and bottom panels are from \textsf{STAGGER} data with the original resolution,
\textsf{STAGGER} data with reduced resolution (HMI PSF),
and \textsf{MURaM} data, respectively. 
The dash-dotted curves indicate the simple Dopplergrams
derived from the Doppler signals made of pairs of filtergrams, namely, 
the core, wing, far-wing, and average-wing Doppler velocities. 
The dashed curves indicate the velocities derived from the filtergrams
(center-of-gravity, or cog, line-center, and the first and second simplified HMI-algorithm 
Dopplergrams),
while the dotted curves indicate the velocities derived from the full spectra
(center of gravity and line center).
 }\label{fig:synth_V_vz_cor_all}.
\end{figure}

Following \inlinecite{2011SoPh..271...27F}, we estimate the 
contribution layer height of  synthetic ``multi-height" Dopplergrams, 
by calculating the correlation coefficients of the line-of-sight velocity derived from 
the synthetic Dopplergrams and the original vertical velocities [$v_z$]
in the simulation box. These correlations are shown in Figure \ref{fig:synth_V_vz_cor_all}.
The heights where the correlation 
coefficients of the Doppler velocities and the vertical velocity 
in the atmosphere attain their maxima are summarized in 
Table \ref{tab:ccmaxheight}. 
For this calculation, we use one snapshot of each simulation dataset.
To estimate the errors in the maximum-correlation heights, 
we subdivide the field of view into
nine areas and calculate the standard deviations of the heights of maximum correlation 
coefficients.
Note that the FOV of \textsf{MURaM} data is 9 Mm square while 
that of \textsf{STAGGER} is 48 Mm square, and this causes larger 
standard deviations of the heights in \textsf{MURaM} datasets. 
In this case the subdivided area in \textsf{MURaM} FOV is about 1 Mm,
namely, about the size of the granular cells. The large standard 
deviation of the height, therefore,  
indicates the spatial variation due to the granular cells.
In Table \ref{tab:ccmaxheight} we also show the results obtained from 
\textsf{CO$^5$BOLD} for reference; these results are from further analysis 
of the datasets used in \inlinecite{2011SoPh..271...27F}.
According to this comparison,
if the resolutions are similar, the contribution-layer heights 
in different atmospheric models are similar.

\begin{figure}[hbt]
\centerline{\hspace*{0.01\textwidth}
\includegraphics[width=0.48\textwidth,clip=]{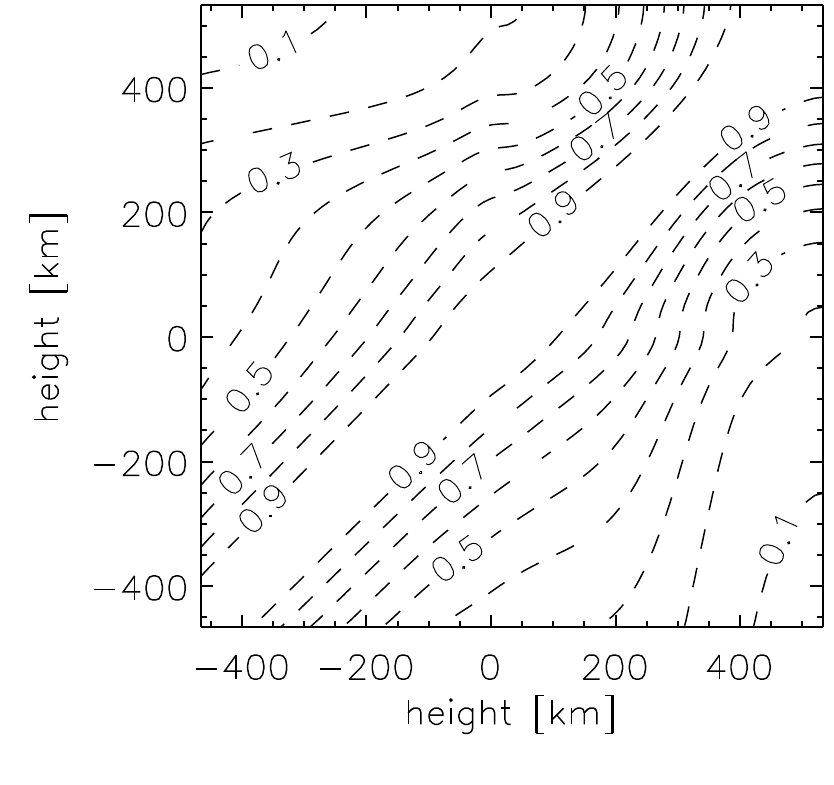}
\hspace*{-0.03\textwidth}
\includegraphics[width=0.48\textwidth,clip=]{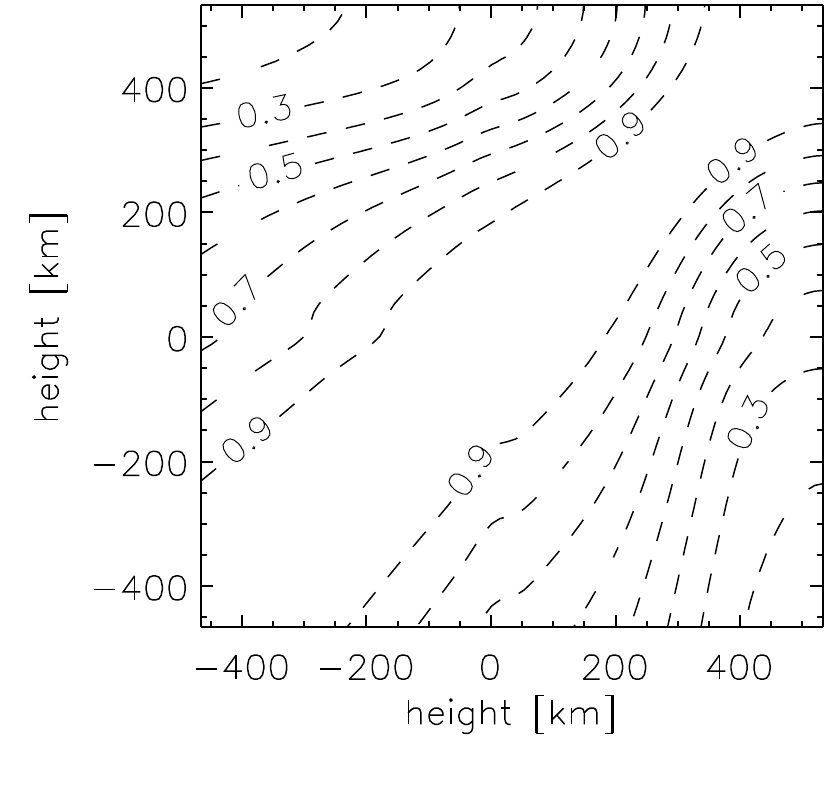}}
\vspace{-0.47\textwidth}
 \centerline{\Large \bf     
      \hspace{0.05\textwidth}   \color{black}{(a)}
      \hspace{0.4\textwidth}  \color{black}{(b)}
         \hfill}
\vspace{0.47\textwidth}
\centerline{\includegraphics[width=0.48\textwidth]{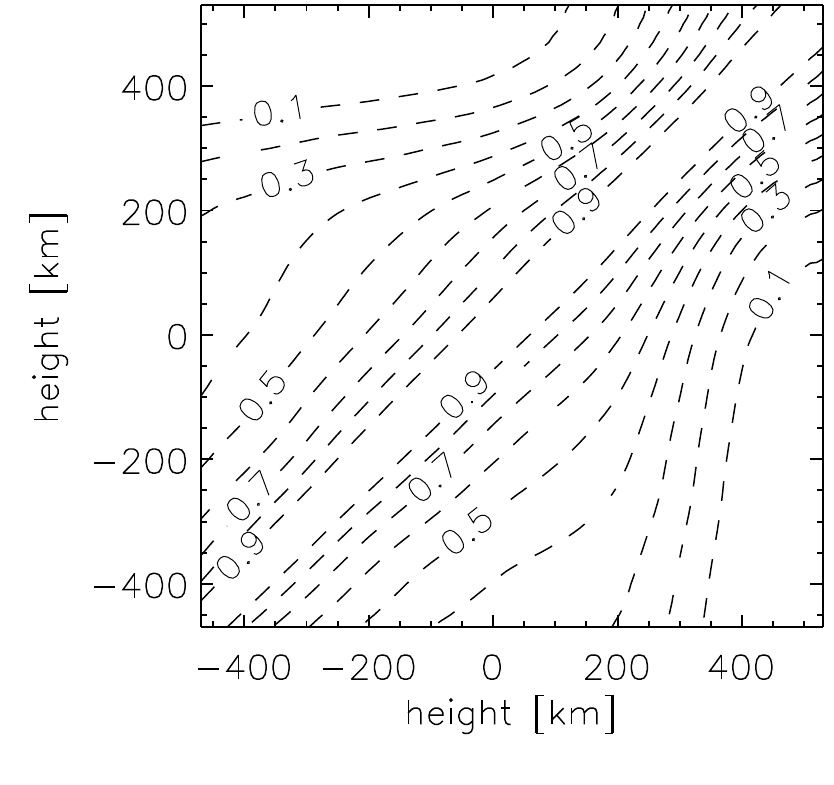}}
\vspace{-0.47\textwidth}
 \centerline{\Large \bf     
      \hspace{0.25\textwidth}   \color{black}{(c)}
         \hfill}
\vspace{0.45\textwidth}
\caption{Autocorrelation coefficients of $v_z$ in the 
atmosphere in the simulation boxes:
\textsf{STAGGER} datacubes with original resolution (a),
\textsf{STAGGER} datacubes with resolution reduced by
the HMI PSF (b), and
\textsf{MURaM} datacube with original resolution (c).} 
\label{fig:syn_Vzcorcoeff}
\end{figure}

Figure \ref{fig:syn_Vzcorcoeff} shows that the auto-correlation coefficient of $v_z$ 
has a broad peak and its FWHM is about 500 km, \textit{i.e.}, 
several scale heights. 
Therefore, this is consistent with what is shown in Figure \ref{fig:synth_V_vz_cor_all};
the Dopplergrams of this wavefield should have a broad range of contribution layers.

\begin{table}[htb]
 \caption{Heights where the correlation coefficient between the 
Doppler velocity and $v_z$ in the atmosphere are maximum. 
The Doppler velocities listed in the titles of the 
rows are defined in Section \ref{sec:DopDef}.
The numbers in the parentheses are the standard deviations of the heights 
of maximum correlation coefficients of nine non-overlapping subareas. 
Note that the FOV of \textsf{MURaM} data is 9 Mm square while that of \textsf{STAGGER} is 48 Mm square. }\label{tab:ccmaxheight}.
\begin{tabular}{lp{6em}p{6em}p{5em}p{5em}}   
\hline
&\multicolumn{4}{c}{Simulated data} \\  
Type of Doppler velocity & \textsf{STAGGER} [km] & 
\textsf{STAGGER} w/ PSF [km]& \textsf{MURaM} [km]  & \textsf{CO$^5$BOLD} [km] \\
\hline
Line center (full) & 234 (6.5) &260 (7.8) & 230 (34) & 230\\
Core & 157 (4.3) & 208 (4.3) & 170 (68) & \\
2nd HMI-algorithm  & 157 (4.3) & 221 (6.5) & 160 (64)&  140 \\
Line Center (3pts) & 144 (0.0) & 221 (8.6) & 150 (68) & 125 \\
1st HMI-algorithm  & 118 (5.7) & 195 (6.8) &110 (56)&100 \\
Center of gravity (full) & 118 (6.8)& 182 (6.8)&  100 (58)& 90 \\
Center of gravity (6pts) & 105 (0.0)& 195 (9.2)& 100 (43)& 90 \\
Wing & 92 (0.0) & 170 (6.5) & 90 (31)& 80\\
Average-wing & 92 (0.0) & 170 (6.5) & 80 (31)& 70 \\
Far-wing & 79 (4.3) & 157 (9.4) & 50 (47) & 55 \\
\hline
\end{tabular}
\end{table}

\inlinecite{2011SoPh..271...27F} show (in their Figure 5) the correlation coefficient between 
the velocity derived from the HMI filtergrams and $v_z$ in their simulation box.
They found that the normal HMI-algorithm Dopplergram and the center of gravity of the line
have a peak correlation with the true $v_z$ around 100 km above the surface, 
while the Doppler shift of the line center has a peak correlation 
around 230 km. These results are consistent with our findings.
Also, \inlinecite{2011SoPh..271...27F} discuss the effects of spatial resolution. 
Their original horizontal grid size is 14 km, and if they apply the estimated 
HMI PSF to their data,
the contribution height increases by about 50 km for their synthetic 
HMI-algorithm Dopplergram. 
On the other hand, our results show 
that the contribution layer heights are  70\,--\,80 km higher if we apply the HMI PSF. 
Since the PSF in \inlinecite{2011SoPh..271...27F} is Gaussian with FWHM of 
1.5 arcsec ($\approx 3$ HMI pixels), the smearing is 
more significant than our PSF
(Equation (\ref{eq:PSF})). 
Why more smearing causes a smaller shift of contribution layer heights is still unclear.
This difference might be caused by other properties, 
such as differences in atmospheric profiles or convection models.

As shown in Table \ref{tab:ccmaxheight} 
the contribution layer of the line-center Dopplergrams
derived from the full spectra is higher than those derived from the filtergrams.
The contribution-layer height is sensitive to the wavelength range
used for the fitting. As we describe in Section \ref{sec:DopDef}, 
we choose $\pm 32$ m\AA \ around the minimum-intensity wavelength,
but if we choose a broader range, the contribution-layer height is reduced.

\subsection{Rather Robust Multi-Height Velocity Datasets} \label{sec:multi-height}

The results of this section tell us that 
we have several ways to derive multi-height velocity information 
in the solar atmosphere from the HMI filtergrams.
We need to consider, however, not only the formation height of each velocity
but also the availability or noise level of the observables.
As was discussed by \inlinecite{2013ASPC..479..429N},
velocities derived from simple Doppler signals [$D_{br}$]
(Doppler velocities 1 in Section \ref{sec:DopDef})
have many unusable points due to the limited wavelength 
separation of the blue and red pair. 
For example, for the core pair the blue and red wings are only
34 m\AA \ away from the line center.
Therefore, if the velocity exceeds 1.7 km  s$^{-1}$, 
the line center is outside of this pair, and
the Doppler signal made from the pair is no longer useful to measure the Doppler velocity. 
Based on the results of this section 
and the availability of observables,
for rather robust multi-height velocity datasets
we propose choosing the average-wing Dopplergrams,
the HMI-algorithm Dopplergrams, and the line-center 
Dopplergrams defined in Section \ref{sec:DopDef}.

\section{Phase Difference} \label{sec:phase}

Using these multi-height Dopplergrams, we calculate the 
phase differences among them.

\subsection{Observed Phase Difference} \label{sec:phase_obs}
For the phase analyses of the HMI observation here, 
the average-wing and the line-center Doppler velocities are calculated
from the six HMI observation filtergrams 
in the same manner as described in Section \ref{sec:DopDef}, while
as for the HMI-algorithm Dopplergrams, we use the standard product of the HMI pipeline
(\opencite{2012SoPh..278..217C}).

\begin{figure}[htbp]
\centerline{\includegraphics[width=0.5\textwidth,clip=]{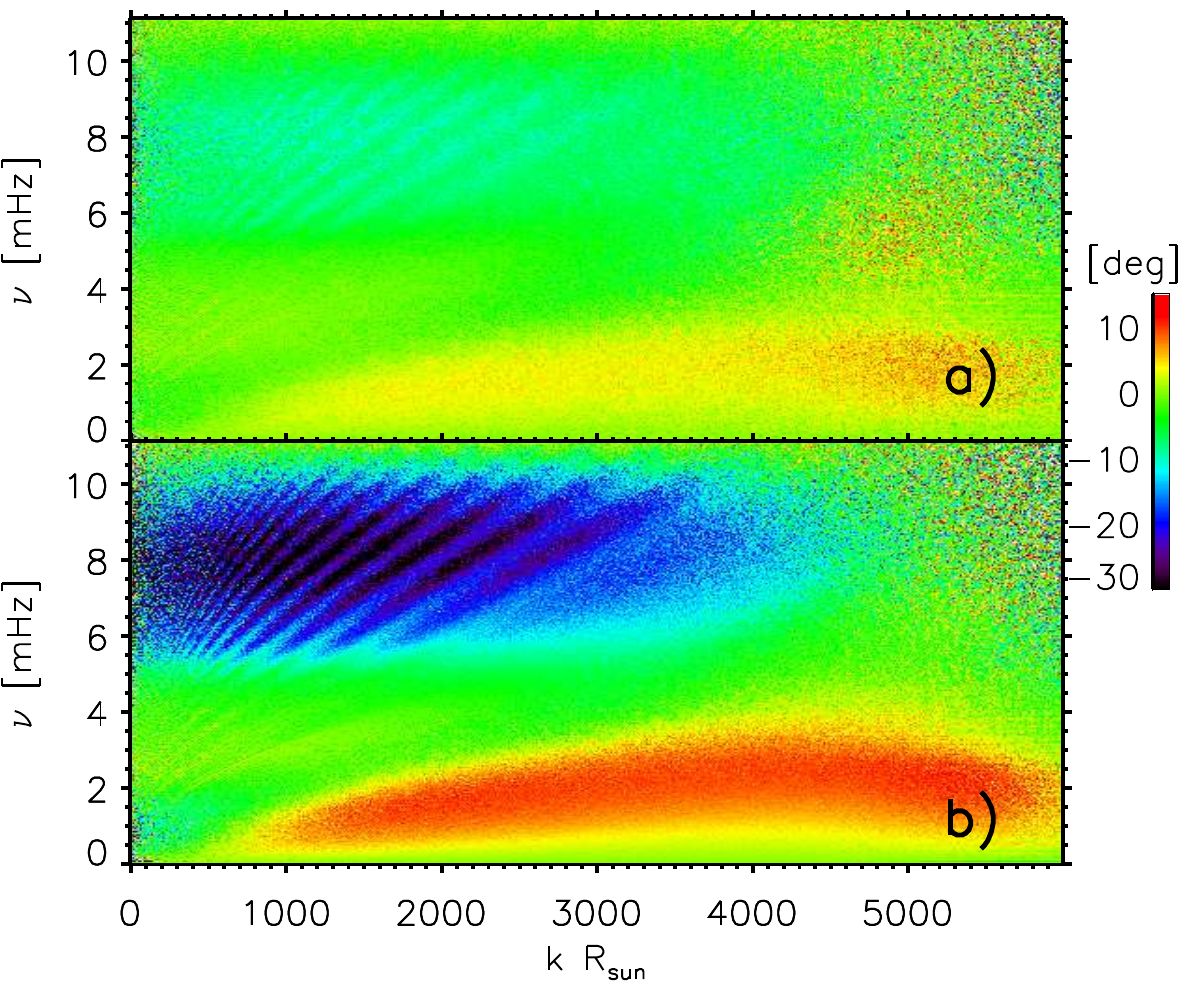}  
 \includegraphics[width=0.416\textwidth,clip=]{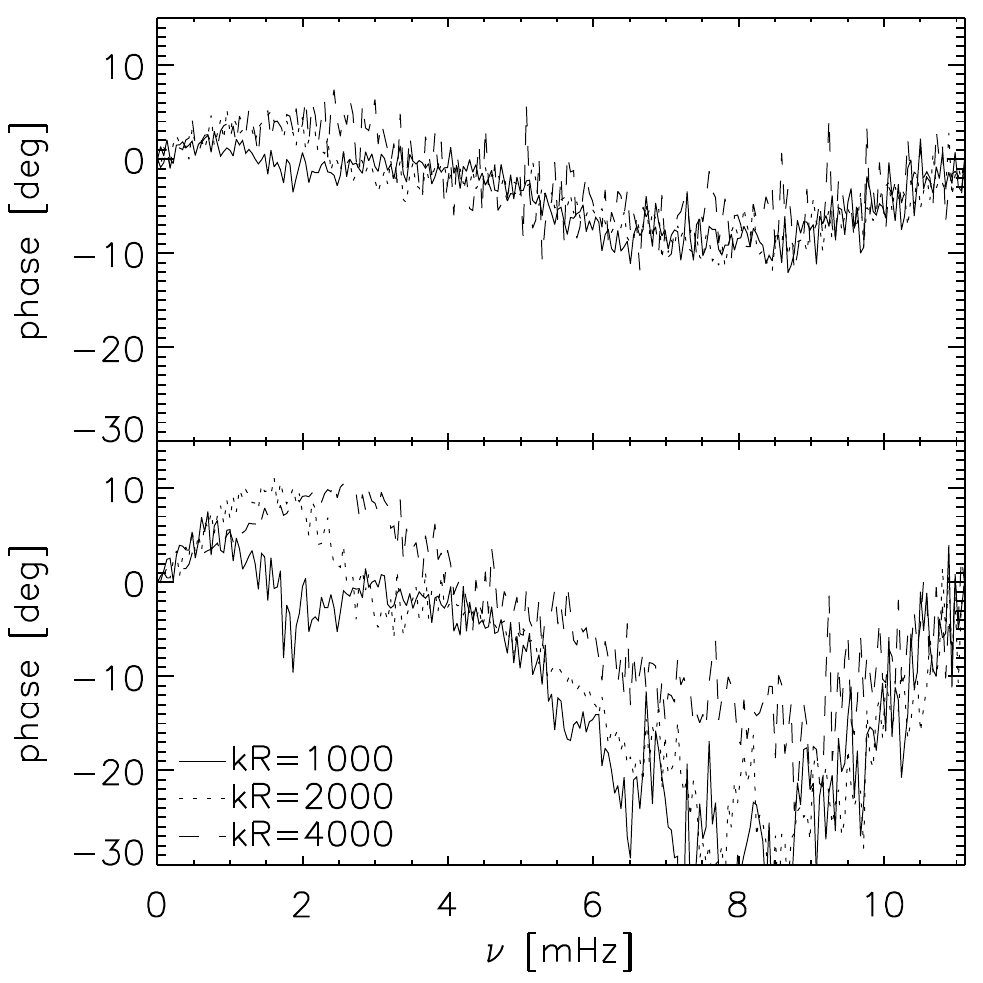}  
}
\centerline{\includegraphics[width=0.5\textwidth]{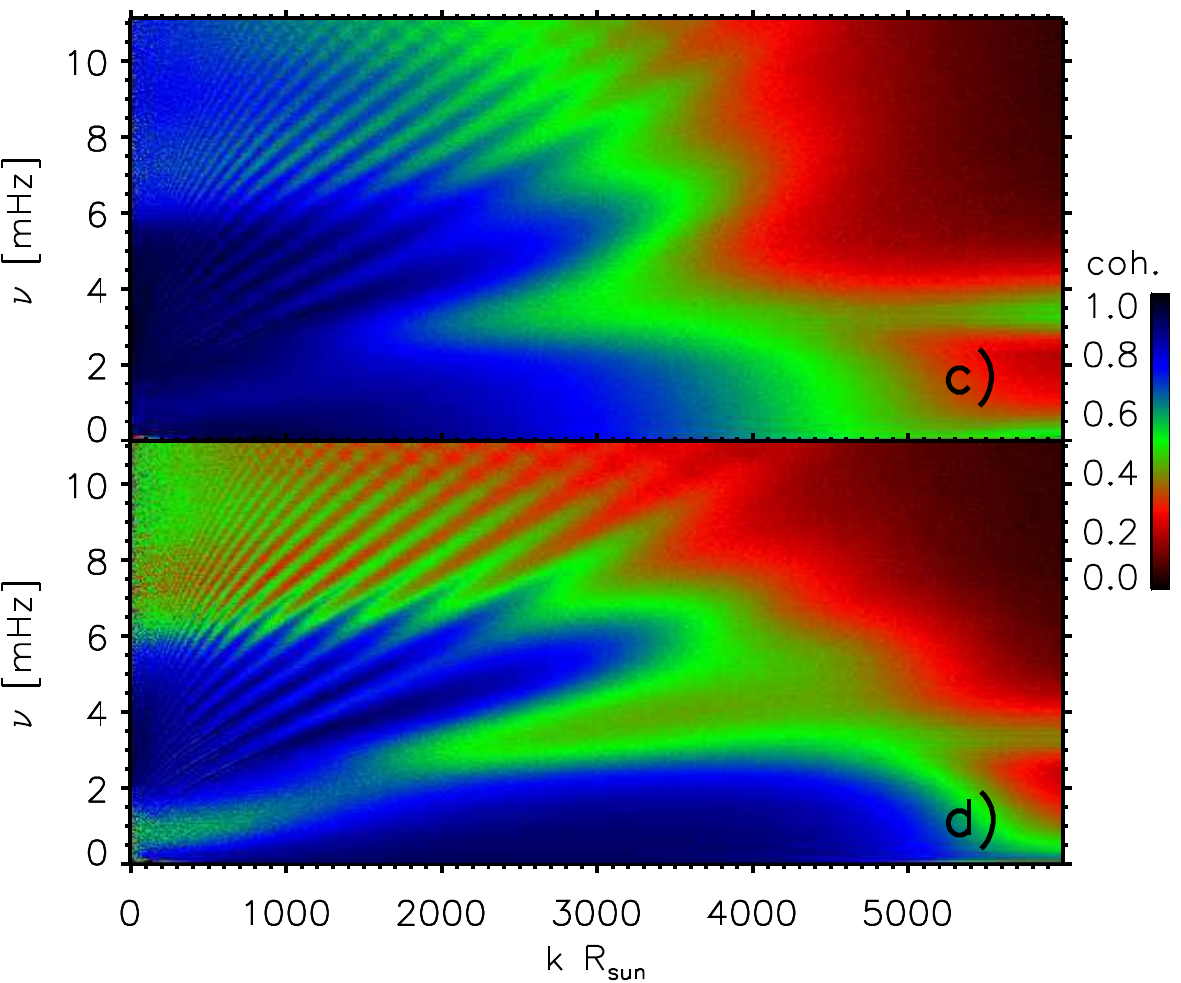}}
\caption{
Top: Observed phase difference between the line-center and  
HMI-algorithm Dopplergrams 
[$\phi_{\mathrm{center}} - \phi_{\mathrm{HMI}}$] (a)
and between the line-center and average-wing Dopplergrams
[$\phi_{ \mathrm{center}} - \phi_{\mathrm{average\mathchar`-wing}}$] (b). 
The horizontal wavenumber [$k$] in the horizontal axis in the left panel
is normalized by the solar radius [$\mathrm{R}_{\odot}$].
Slices at $k \mathrm{R}_{\odot}=1000, 2000, 4000$ are shown 
in the top--right panel.
Bottom: coherence spectra of the line-center and
HMI-algorithm Dopplergrams (c), and
the line-center and average-wing 
Dopplergrams (d).
}\label{fig:obs_phasedif} 
\end{figure}

Given Fourier transform 
$\tilde{f}(\vec{k}, \omega) \equiv |\tilde{f}| e^{\mathrm{i}\phi_f}$ 
and $\tilde{g}(\vec{k}, \omega) \equiv |\tilde{g}| e^{\mathrm{i}\phi_g}$
of two Doppler velocities, $f(\vec{x}, t)$ and $g(\vec{x}, t)$,
where $\vec{k}=(k_x, k_y)$ is the spatial wavenumber, $\omega$ is 
the temporal wavenumber, and $\vec{x}=(x,y)$ and $t$ are the location and the time, 
respectively, and $\phi_f, \phi_g$ are real,
we define the cross spectrum as
$\tilde{f}(\vec{k}, \omega) \tilde{g}^*(\vec{k}, \omega) 
=|\tilde{f}||\tilde{g}| e^{\mathrm{i}(\phi_f -\phi_g)}$.
For Figure \ref{fig:obs_phasedif} we azimuthally average 
$\tilde{f} \tilde{g}^*$ in the $k_x$-$k_y$ plane, 
and calculate the phase of the average as a function of 
horizontal wavenumber [$k \equiv ||\vec{k} ||$] 
and frequency [$\nu=\omega/(2\pi)$].
In what follows the phase of the azimuthally averaged $\tilde{f} \tilde{g}^*$
is denoted by simply $\phi_f-\phi_g$.
This $\phi_f-\phi_g$ ranges from $-\pi$ to $+\pi$.
We also calculate the coherence to show how reliable the phase
has been determined. Here we define the coherence of the two 
velocity fields [$f$ and $g$] as 
$|\langle\tilde{f} \tilde{g}^*\rangle|/(\langle |\tilde{f}|^2\rangle 
\langle |\tilde{g} |^2\rangle)^{1/2}$, where $\langle \rangle$ means 
azimuthal average at each $k$.

Figure \ref{fig:obs_phasedif} shows the phase differences between the 
average-wing, the HMI-algorithm, and the line-center Dopplergrams,
as well as their coherences. 
For this calculation, we use 512 frames with 45-second cadence (384 minutes) 
starting from 4:17:15 UT on 23 January 2011.
The region is a 30-square-degree quiet region near the disk center.
This region is tracked at the Carrington rate by using \textsf{mtrack}
(\opencite{2011JPhCS.271a2008B}),
and the central point of the area passes the heliographic disk center 
at the mid-point of the time period. 
\textsf{Mtrack} is a module used in the HMI pipeline to make tracked and mapped datacubes at 
selected heliographic coordinates.

The phase difference shows three characteristic features
in $k$-$\omega$ space as we expect.
First, in the $p$-mode regime (lower frequency, lower wavenumber),
the phase difference is nearly zero
because they are eigenmodes of the Sun.
Second, above the acoustic-cutoff frequency in the photosphere 
($\approx$ 5.4 mHz) 
the phase difference is negative and proportional to the frequency
which indicates upward-propagating waves.
Third, in the convective regime, namely, in the lower frequency and the larger wavenumber
ranges, the phase difference is positive.
This is a signature of atmospheric gravity waves, which was reported,
\textit{e.g.}, by \inlinecite{2008ApJ...681L.125S} for \textsf{CO$^5$BOLD} 
datasets and by \inlinecite{2009ASPC..415...95S} in observations.
Note that in the highest frequency range ($\nu>8$ mHz) 
the phase difference
cannot be trusted because of their proximity to the Nyquist frequency and
possible aliasing effects.
The coherence spectra show that these characteristic phase-difference features
are reliable over most of $k$-$\omega$ space,
in particular in the $p$-mode and internal-gravity wave regimes.
At high wave numbers ($kR_{\odot} \gtrsim 4000$)
and high frequencies (red areas in the coherence diagrams),
the measured phase signal becomes unreliable
(also note the scatter of the phase signal in this area
in the $k$-$\omega$ phase diagrams).
It is beyond the scope of this article to discuss the differences
between the pseudo $p$-mode ridges and the surrounding inter-ridges,
which remain a puzzle,
in both the phase-difference and coherence spectra.

The phase differences
above the acoustic-cutoff frequency in the photosphere ($\approx$ 5.4 mHz) 
increase linearly as the frequency increases.
Around 8 mHz, the phase differences [$\Delta \phi$] are 
\begin{eqnarray}
\Delta \phi_{\mathrm{center} - \mathrm{HMI}} &=&
\phi_{\mathrm{center}} - \phi_{\mathrm{HMI}} \approx -10^{\circ} \\
\Delta \phi_{\mathrm{center} - \mathrm{average\mathchar`-wing}} &=&
\phi_{\mathrm{center}} - \phi_{\mathrm{average\mathchar`-wing}} \approx -30^{\circ} . 
\end{eqnarray}
This minus sign of the phase means upward propagation.
The phase of the line-center Dopplergram is smaller,
and, therefore, the layer indicated by the line-center 
Dopplergram is higher than
the layers indicated by the HMI-algorithm Dopplergram 
and average-wing Dopplergram.
If we simply estimate the height difference between the two formation layers
[$\Delta z$] from $\Delta \phi$ by 
$\Delta z/c_\mathrm{s} = -\Delta \phi/(2 \pi \nu)$,
where $c_\mathrm{s} \approx 7 \ \mathrm{km \ s}^{-1}$ is the photospheric sound speed, we obtain
\begin{eqnarray}
\Delta z_{\mathrm{center-HMI}}&=& -c_\mathrm{s}
 \Delta \phi_{\mathrm{center} - \mathrm{HMI}} /(2 \pi \nu)
\approx 24 \ \mathrm{km} \\
\Delta z_{\mathrm{center-average\mathchar`-wing}}&=&  -c_\mathrm{s} 
\Delta \phi_{\mathrm{center} - \mathrm{average\mathchar`-wing}}/(2 \pi \nu)
\approx 73 \ \mathrm{km}.
\end{eqnarray}
The height difference $\Delta z_{\mathrm{center-HMI}}\approx 24 \ \mathrm{km}$
is similar to the height difference between the estimated layers
where the velocity signal based on the 
line-center (three points) and the first HMI-algorithm Dopplergrams form 
(26 km, see Table \ref{tab:ccmaxheight}), while 
$\Delta z_{\mathrm{center-average\mathchar`-wing}} \approx 73 \ \mathrm{km}$ corresponds to
the estimated height difference between the layers
where the velocity signal based on the line-center (three points) and 
average-wing Dopplergrams form (52 km, see Table \ref{tab:ccmaxheight}).
However, they are not necessarily identical to each other,
because the phase difference is measured 
at a particular wave mode  $(k,\omega)$,
while the height measured by taking the correlation coefficients in 
Section \ref{sec:syn_cor} shown in Table \ref{tab:ccmaxheight}
is for the total velocity field.
This is further discussed in the next subsection.

\subsection{Velocity Response Function}
The vertical velocity discussed in Section \ref{sec:syn_cor}
is a superposition of turbulent-convective and wave motions, and 
the height at which the correlation between the velocity and Doppler 
signal is highest is based on both the wave and convective motions. 
In this section we discuss the velocity response functions. 
If the velocities due to the waves were small 
compared to the convective motions, then 
the height at which the Doppler signal is affected
by the wave motions is determined by the response function
 (and the vertical component of the wave velocities). 
If the velocities due to the waves turn out to be substantial, 
the response functions therefore do not allow the height range 
to be cleanly determined where the Doppler signal is 
influenced by the waves. 
None the less, the response functions give some insight into the problem.

We computed the velocity response function by 
\textsf{a program to compute full STOkes PROfiles of Zeeman split atomic and
molecular absorption lines in LTE} (\textsf{STOPRO}) included in 
the \textsf{SPINOR} code (\opencite{2000A&A...358.1109F}).
The response function [$R(\lambda, \tau, v_z)$] is the function 
indicating how the vertical-velocity field perturbation in each layer 
 (in terms of optical depth [$\tau$]) affects the intensity, and is defined by
\begin{eqnarray}
I(\lambda, v_z') -I(\lambda, v_z)=
I_{\mbox{cont}}\int \mathrm{d}\tau R(\tau, \lambda, v_z) [v_z'(\tau)-v_z(\tau)] \ ,
\end{eqnarray}
where $I(\lambda,v_z)$ is the intensity at the wavelength $\lambda$
if the vertical velocity field is $v_z=v_z (\tau)$,
$v_z'(\tau)$ is the perturbed vertical velocity,
and $I_{\mbox{cont}}$ is the continuum intensity.
The response function thus gives the linear sensitivity of the intensity
to a small change in the velocity field.

\begin{figure}
\centerline{
\includegraphics[width=0.8\textwidth]{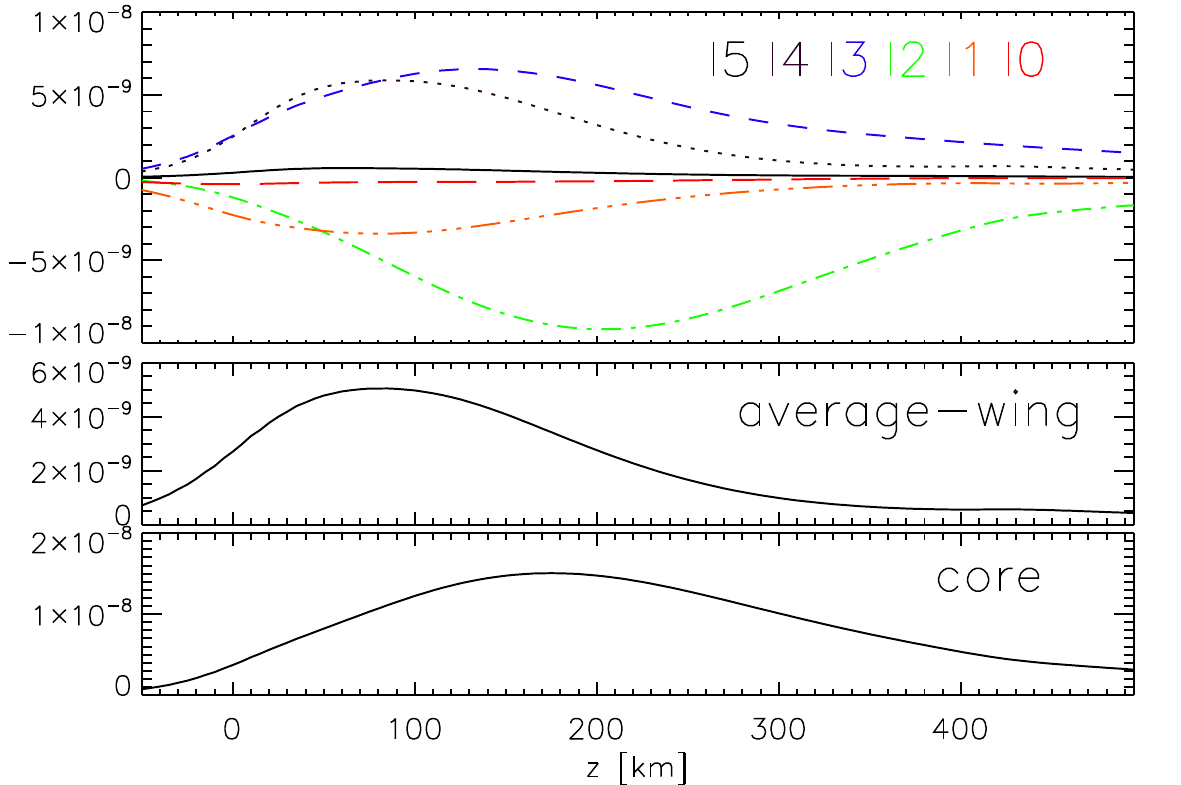} }
\caption{Velocity response functions convolved with the HMI filter profiles 
(shown in Figure \ref{fig:prof_line}) made from 
\textsf{STAGGER} simulation dataset. The functions are calculated
by the \textsf{SPINOR} code for each pixel and then averaged over an area with 10 Mm square. 
The units of the response functions are (cm s$^{-1}$ km)$^{-1}$. 
The top panel shows the response functions for the filtergrams 
(I0: red long-dashed, I1: orange dash-triple-dotted, I2: green dash-dotted,
I3: blue short-dashed, I4: dark-purple dotted, I5: black solid),
while the bottom two panels show 
the response functions for the  average-wing and
core Doppler velocities.
} \label{fig:resp}
\end{figure}

Figure \ref{fig:resp} shows the response functions.
The function is calculated at each pixel in a snapshot and averaged over 
a 10-Mm-square area of \textsf{STAGGER} datacubes convolved with the HMI filter profiles.
To make this figure, using the conversion of the 
optical depth [$\tau$] into the geometrical height [$z$], we plot 
$R(\tau,\lambda,v_z) \mathrm{d}\tau /\mathrm{d}z$ as a function of $z$. 
From Figure \ref{fig:resp}, it is clear that the sensitivity of the 
far-wing pair [$I_0$ and $I_5$] 
to the velocity is low, and the far-wing Dopplergrams might not be reliable.

For simplicity, here we consider the simple Dopplergram cases only (Doppler velocities 1 
defined in Section \ref{sec:DopDef}) using these response functions.
If we assume the denominator of 
$D_\mathrm{br} = (I_\mathrm{b} -I_\mathrm{r})/(I_\mathrm{b}+I_\mathrm{r}) $
does not have much dependence on the velocity,
the response function for $D_\mathrm{br}$ is
$R_\mathrm{br} \approx R(\tau,\lambda_\mathrm{b}, v_z) 
- R(\tau,\lambda_\mathrm{r}, v_z)$.
These $R_\mathrm{br}$ are shown in Figure \ref{fig:resp} as well.
The heights of the center of gravity of $R_\mathrm{br}$ are
140 km for average-wing and 210 km for core, and the formation height
difference between average-wing and the core is 70 km.
Although this simple core Doppler velocity is not identical to the 
line-center Doppler velocity, here we may use the core Doppler velocity as a proxy for the
line center. Then, this is consistent with the rough estimate of the height difference
derived from the earlier phase-difference measurement,
$\Delta z_{\mathrm{center-average\mathchar`-wing}} \approx 73 \ \mbox{km}$.

\subsection{Simulated Phase Difference}
\begin{figure}[htbp]
{\includegraphics[width=0.5\textwidth,clip=]{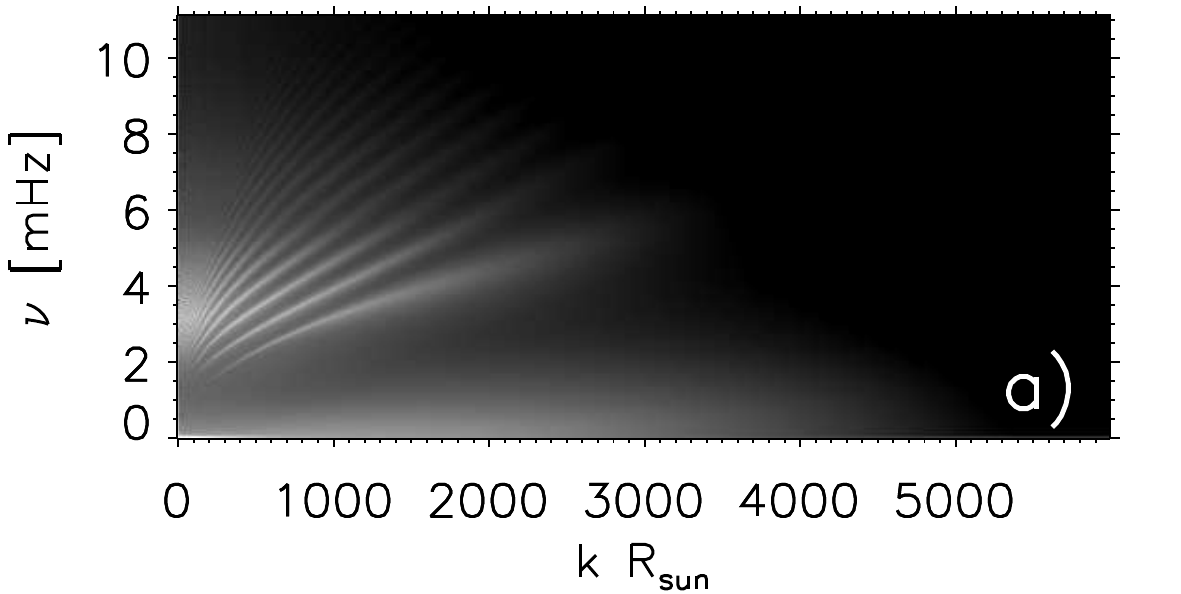}  
\hspace{-0.05\textwidth}
\includegraphics[width=0.5\textwidth,clip=]{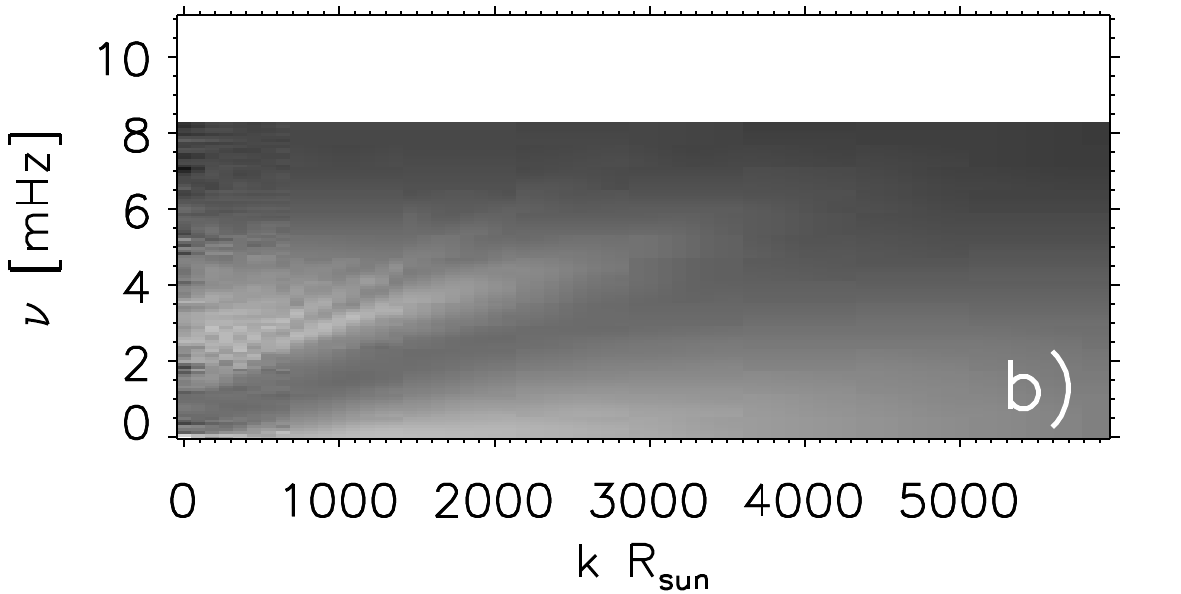} }
\centerline{\includegraphics[width=0.8\textwidth]{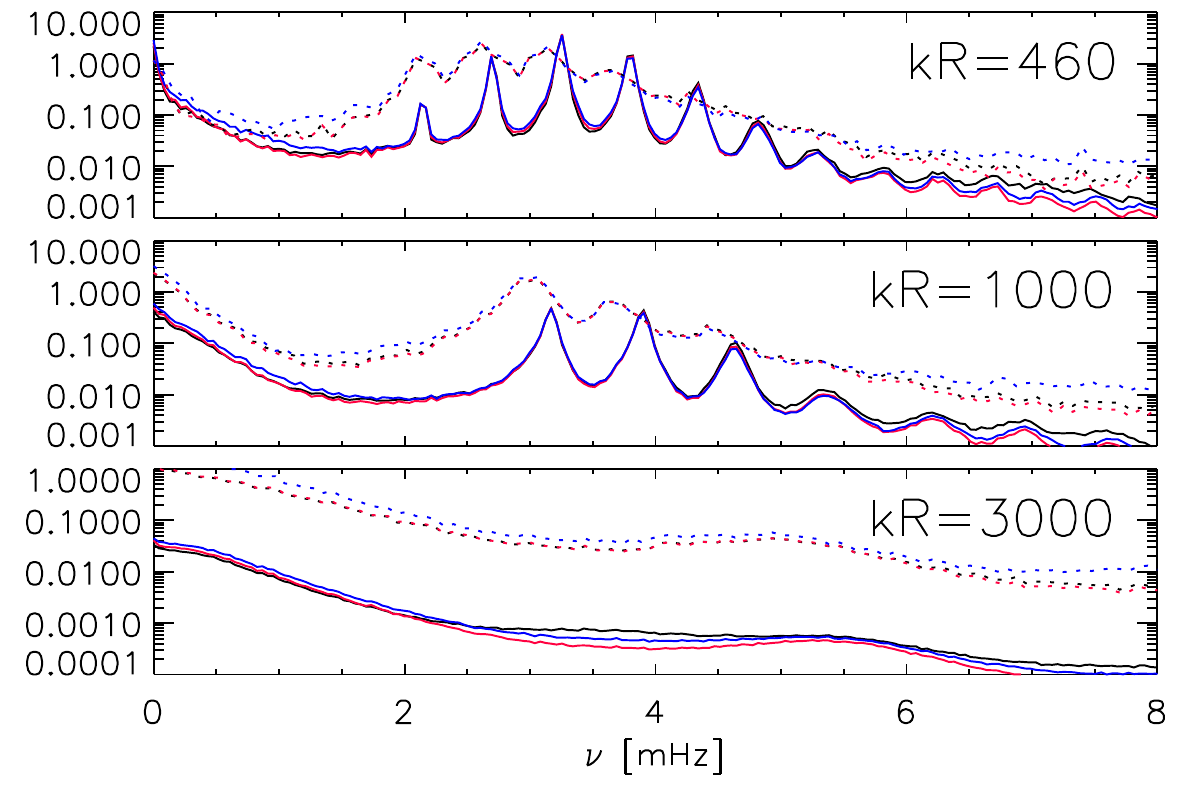}
}
\caption{Power spectra of the HMI-algorithm Dopplergrams made from 
HMI observation data (a) and
from \textsf{STAGGER} simulation data (b) 
in logarithmic gray scaling (white indicates higher power, 
while black indicates lowest power).
The observation and simulated datasets are identical to what we use for
Figures \ref{fig:obs_phasedif} and \ref{fig:syn_phasedif}, respectively. 
Note that the Nyquist frequency of the simulation data (b) is 
lower than that of the observation (a),
because of the difference in cadence 
($\Delta t = 60 \ \mathrm{seconds} $ instead of 45 seconds).
The bottom panel shows slices at several $k\mathrm{R}_{\odot}$. 
The solid and dotted curves indicate the observations and \textsf{STAGGER} 
data. Line center, HMI-algorithm Dopplergram, and average-wing Doppler 
velocities are in black, red, and blue, respectively. Each power
spectrum is normalized by its average power 
in the region $100 \le k\mathrm{R}_{\odot} \le 500$ and 
$3 \mathrm{mHz} \le \nu \le 3.5 \mathrm{mHz}$, 
and in the power map the grayscale range is from  $10^{-4}$ to $10^2$.
}\label{fig:obs_syn_power} 
\end{figure}

To compare with the observed phase differences,
we calculate the phase differences in several ways using the \textsf{STAGGER}
simulation. The oscillation power in the HMI observations and 
the \textsf{STAGGER} simulation are 
shown in Figure \ref{fig:obs_syn_power}. Although the power contrast of the 
$p$-mode peaks is weaker in the \textsf{STAGGER} simulation, the general trends agree with 
each other. In particular, both spectra have power humps 
in the convection/internal-gravity wave regime, 
namely, in the lower-frequency range.

We calculate the phase difference of the 221-minute-long 
\textsf{STAGGER} simulation data 
time series with one-minute cadence in the following ways.

\subsubsection{Phase Difference 1: Dopplergrams from Synthesized Filtergrams}
\begin{figure}[htbp]
{\includegraphics[width=0.5\textwidth,clip=]{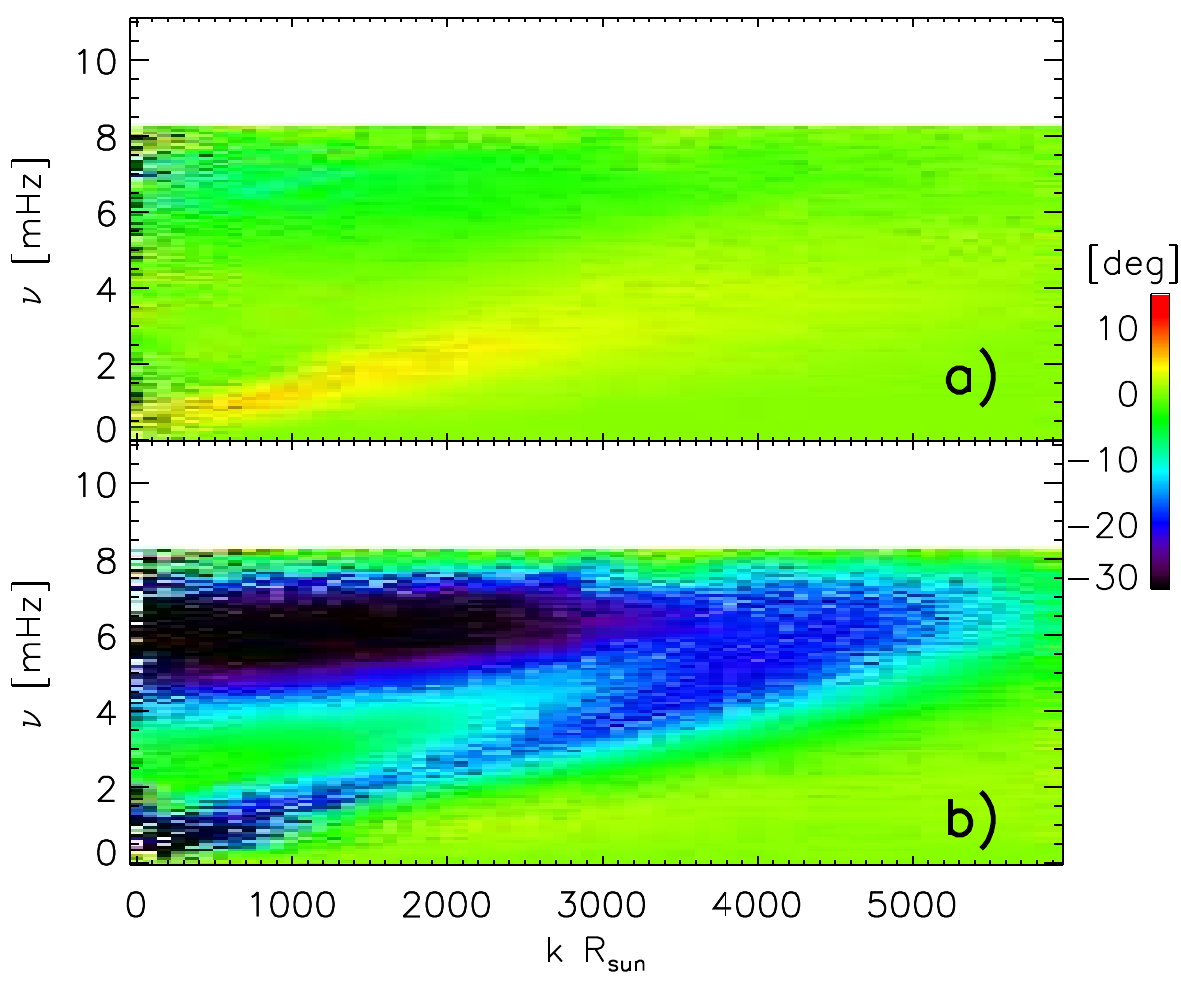}  
\includegraphics[width=0.416\textwidth,clip=]
{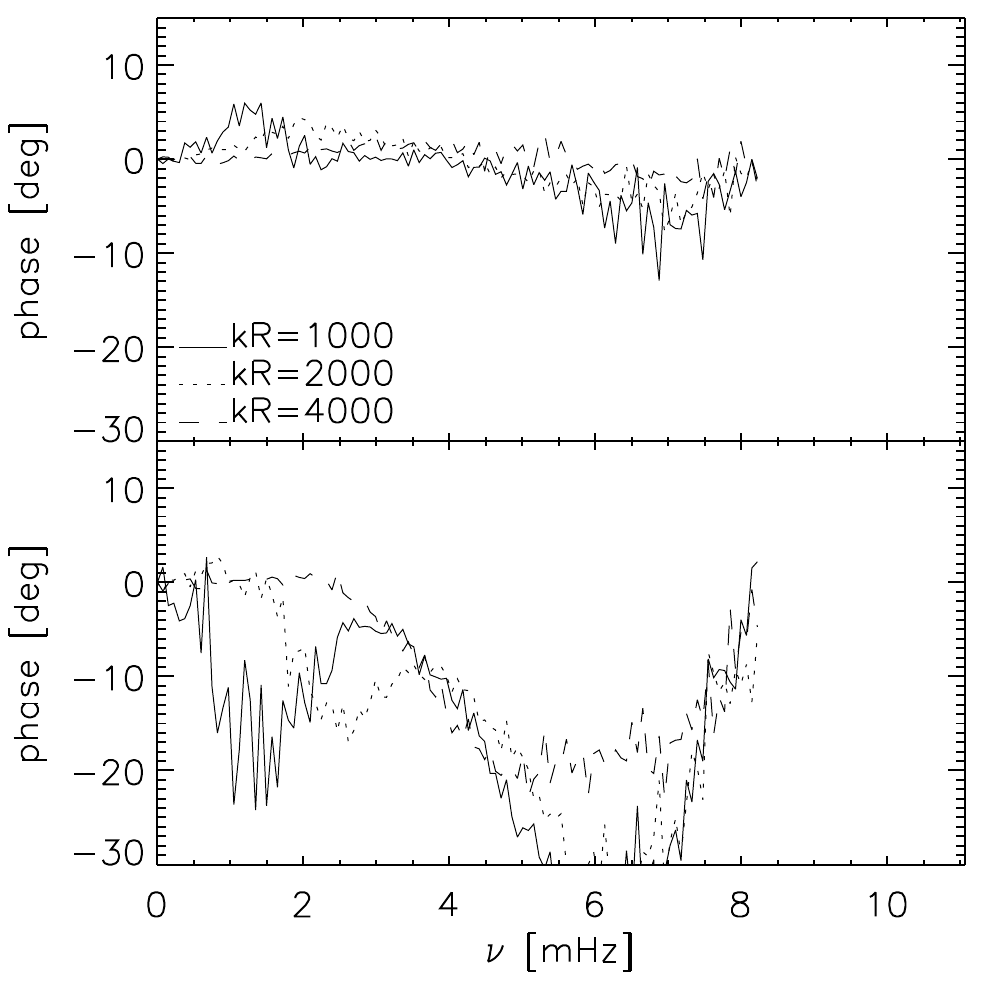} }
\caption{Phase difference between the line-center and 
first HMI-algorithm Dopplergram, 
$\phi_{\mathrm{center}} - \phi_{\mathrm{HMI}}$ (a)
and between the line-center and average-wing 
$\phi_{ \mathrm{center}} - \phi_{\mathrm{average\mathchar`-wing}}$ (b),
from \textsf{STAGGER} simulation data.
Slices at $k \mathrm{R}_{\odot}=1000$, $2000$, and $4000$ are shown 
in the right panel.
 }\label{fig:syn_phasedif} 
\end{figure}

We use the same line-center, HMI-algorithm, 
and the average-wing Doppler velocities
as those used in the observation data plot (Figure \ref{fig:obs_phasedif}).
The phase differences between them are shown in Figure \ref{fig:syn_phasedif}.

\subsubsection{Phase Difference 2: Dopplergrams from Synthesized Spectra}
\begin{figure}[htbp]
{\includegraphics[width=0.5\textwidth,clip=]{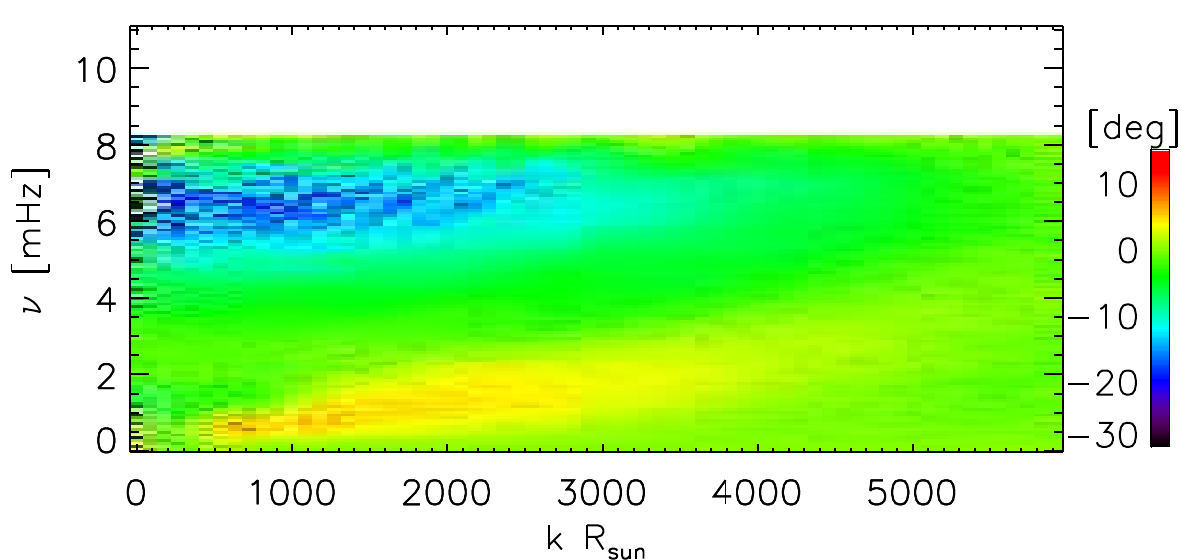}  
\includegraphics[width=0.416\textwidth,clip=]
{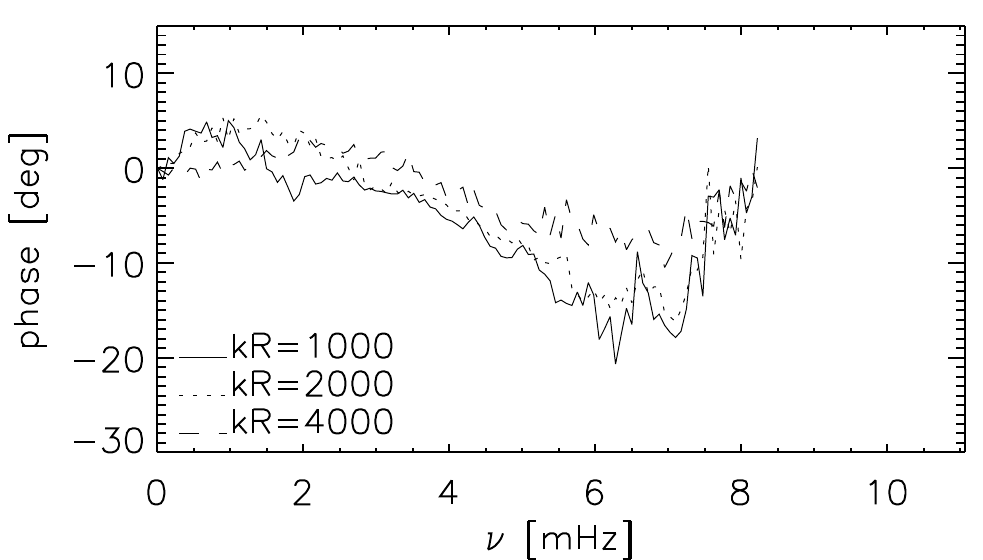} }
\caption{Phase difference between the line-center and center of gravity Dopplergrams
$\phi_{\mathrm{center, full}} - \phi_{\mathrm{cog, full}}$
synthesized from \textsf{STAGGER} simulation data.
Note that these are calculated not from the synthesized filtergrams but from the full spectra,
namely they are Dopplergrams 5 and 6 defined in Section \ref{sec:DopDef}.
Slices at $k \mathrm{R}_{\odot}=1000$, $2000$, and $4000$ are shown in the right panel. 
The contribution height of the center-of-gravity Dopplergram is very 
close to the first HMI-algorithm Dopplergram, and
this figure is a counterpart of Figure \ref{fig:syn_phasedif}a,
although the contribution layer of line-center Dopplergram made from the full spectrum is
higher than that made from filtergram, which makes the phase difference larger than 
 Figure \ref{fig:syn_phasedif}a.
 }\label{fig:syn_phasedif_fullspec} 
\end{figure}

Although only filtergrams rather than full spectra 
are obtained from the HMI observations,
here for comparison
we use the center-of-gravity and line-center Doppler velocities
calculated from the full synthetic spectra,
which are defined in Section \ref{sec:DopDef}.
The phase difference between them is shown 
in Figure \ref{fig:syn_phasedif_fullspec}.
According to the results in Section \ref{sec:syn_cor} 
and those of \inlinecite{2011SoPh..271...27F}, 
the contribution height of the center-of-gravity Dopplergram is 
very similar to the first HMI-algorithm Dopplergram, although,
according to Table \ref{tab:ccmaxheight},
the contribution heights of the line-center Dopplergram 
made from the full spectrum is much higher than that made from 
filtergrams. Therefore, this figure is a counterpart of 
Figure \ref{fig:syn_phasedif}a.
It is not a surprise that
the absolute values of phase difference both in the 
acoustic regime (propagative acoustic waves) and 
in the convection regime (atmospheric gravity waves)  
are larger than those in Figure \ref{fig:syn_phasedif}a;
This is because the height difference between the Dopplergrams
made from the full spectrum is larger
(see Table \ref{tab:ccmaxheight}). 

\subsubsection{Phase Difference 3: Vertical Velocities at Iso-Optical-Depth Surfaces}
\begin{figure}[htbp]
{\includegraphics[width=0.5\textwidth,clip=]{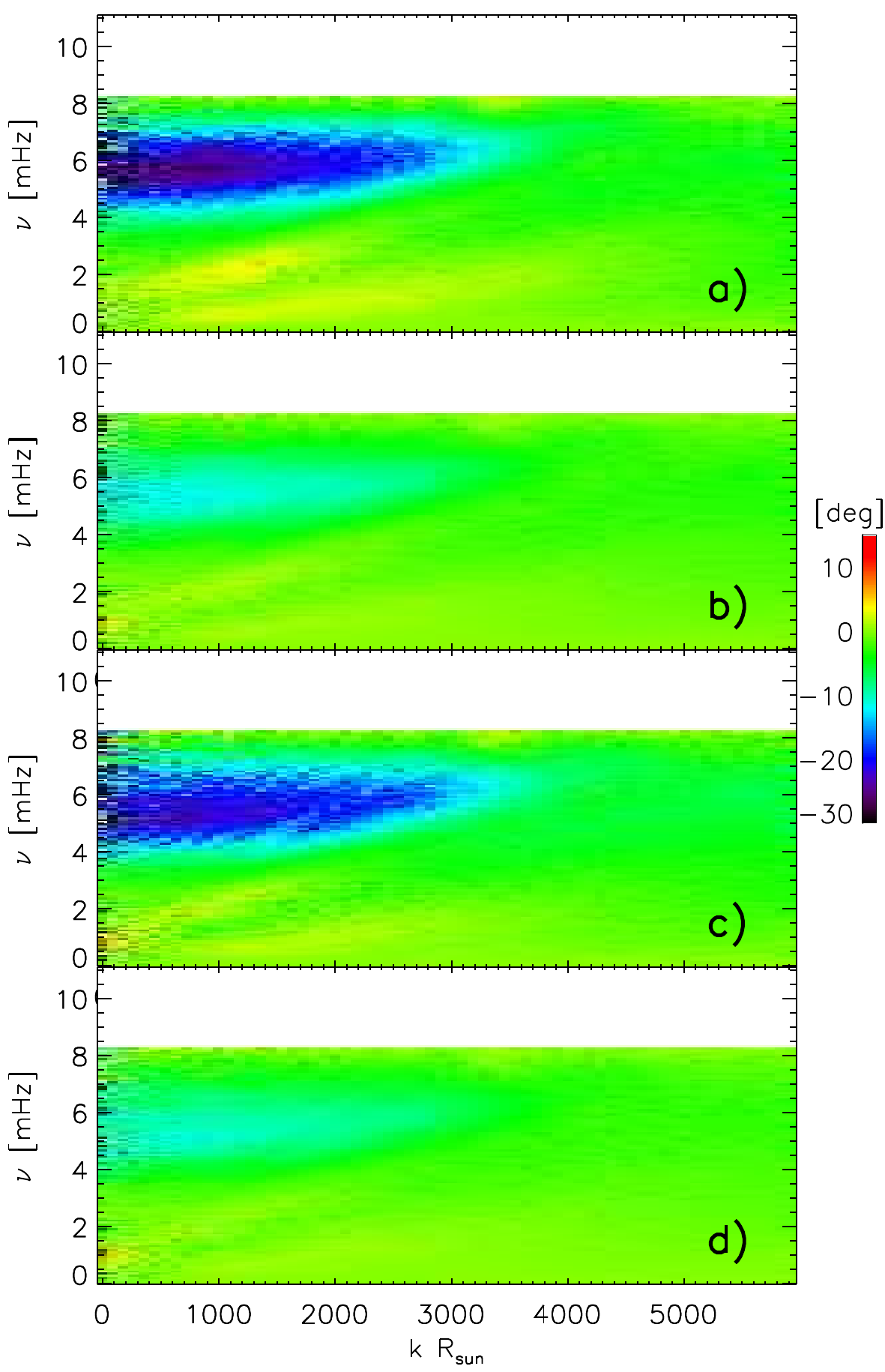}  
 \includegraphics[width=0.416\textwidth,clip=]{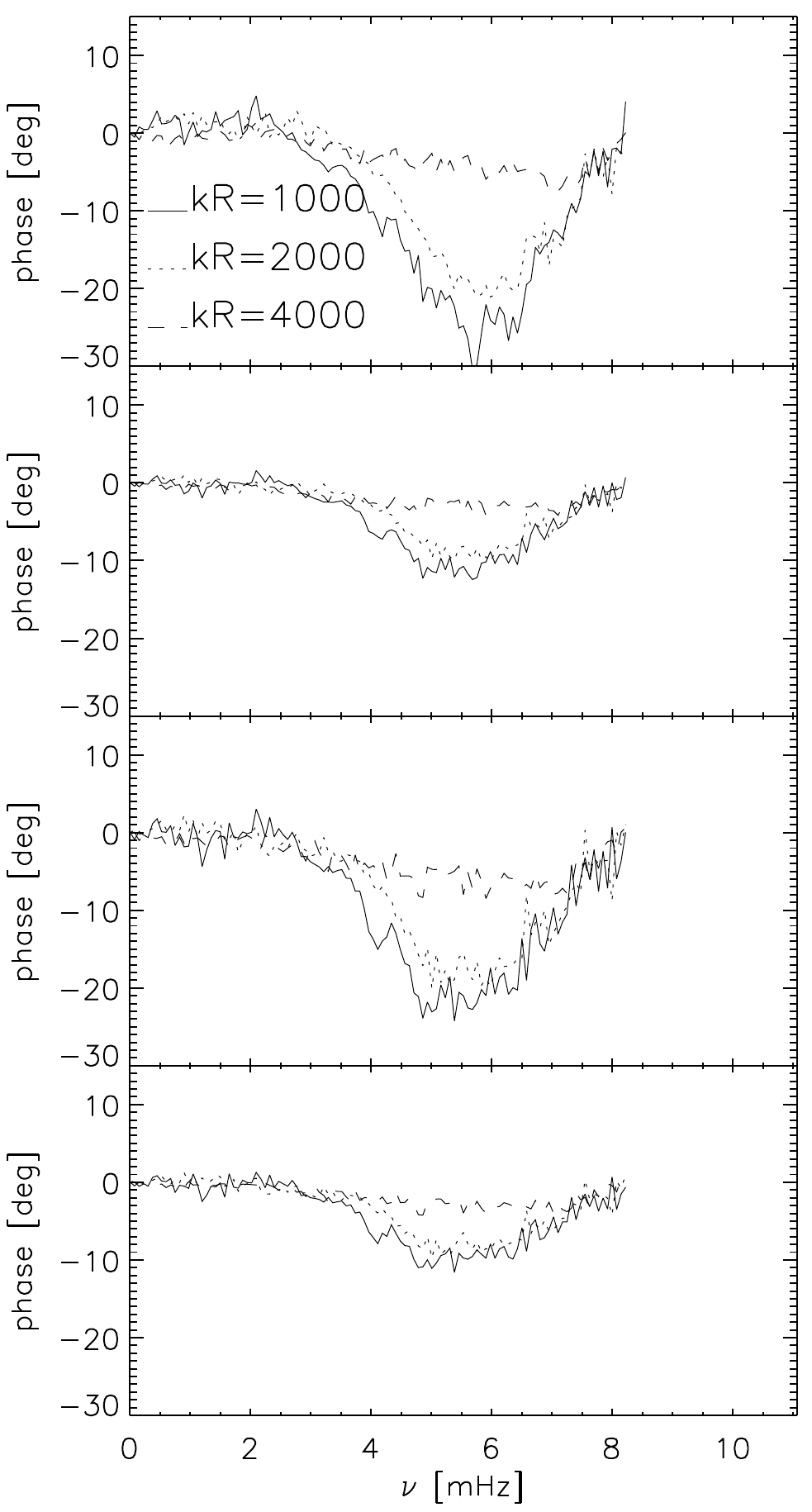} }
\caption{Phase difference of the vertical velocity [$v_z$] between 
pairs of iso-optical depth surfaces in \textsf{STAGGER} simulation atmosphere:
(a) $\phi (\log \tau \!\! =\!\!-1.5) - \phi (\log \tau \!\! =\!\!-1)$,
(b) $\phi (\log \tau \!\! =\!\!-1) - \phi (\log \tau \!\! =\!\!-0.75)$, 
(c) $\phi (\log \tau \!\! =\!\!-1) -\phi (\log \tau \!\! =\!\!-0.5)$, and 
(d) $\phi (\log \tau \!\! =\!\!-0.75)-\phi (\log \tau \!\! =\!\!-0.5)$. 
The slices at $k \mathrm{R}_{\odot}=1000$, $2000$, and $4000$ are shown in the right panel.
The geometrical heights calculated in the atmosphere averaged over the field of view are
223\, km ($\log \tau \!=\!-1.5$), 
145\, km ($\log \tau \!=\!-1$),
104\, km ($\log \tau \!=\!-0.75$), and 62 km ($\log \tau \!=\!-0.5$),
although the relationship between the geometrical height and optical depth 
significantly varies from point to point.
}\label{fig:syn_phasedif_Vz_isotau} 
\end{figure}

We choose the layers with the optical depth $\log \tau \! = (-1.5, -1, -0.75,-0.5)$,
and calculate the phase differences of the vertical velocity field [$v_z$]
among these layers.
It is not straightforward to estimate the geometrical heights of these layers,
because the relationship between the geometrical heights and optical depths 
significantly varies from point to point.
To get a very rough idea, however, we here show the geometrical heights calculated 
in the average \textsf{STAGGER} atmosphere:
223 km ($\log \tau \!=\!-1.5$), 
145 km ($\log \tau \!=\!-1$), 104 km ($\log \tau \!=\!-0.75$)
and 62 km ($\log \tau \!=\!-0.5$) above the surface ($\log \tau=0$).
Figure \ref{fig:syn_phasedif_Vz_isotau} shows the phase differences 
among these layers.

\subsubsection{Phase Difference 4: Vertical Velocities at Iso-Geometrical 
Height Layers}
\begin{figure}[hbtp]
{\includegraphics[width=0.5\textwidth,clip=]{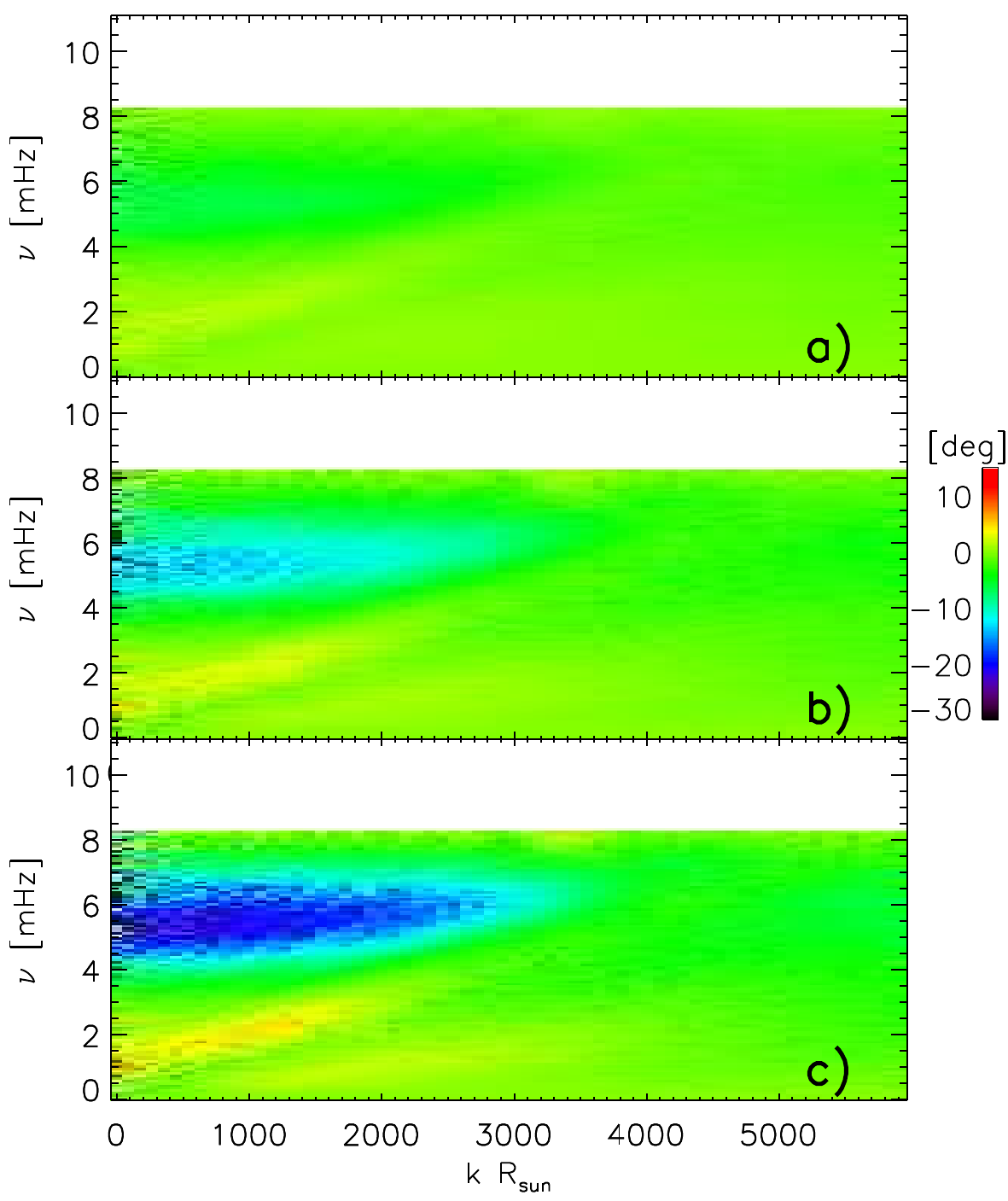}  
 \includegraphics[width=0.416\textwidth,clip=]{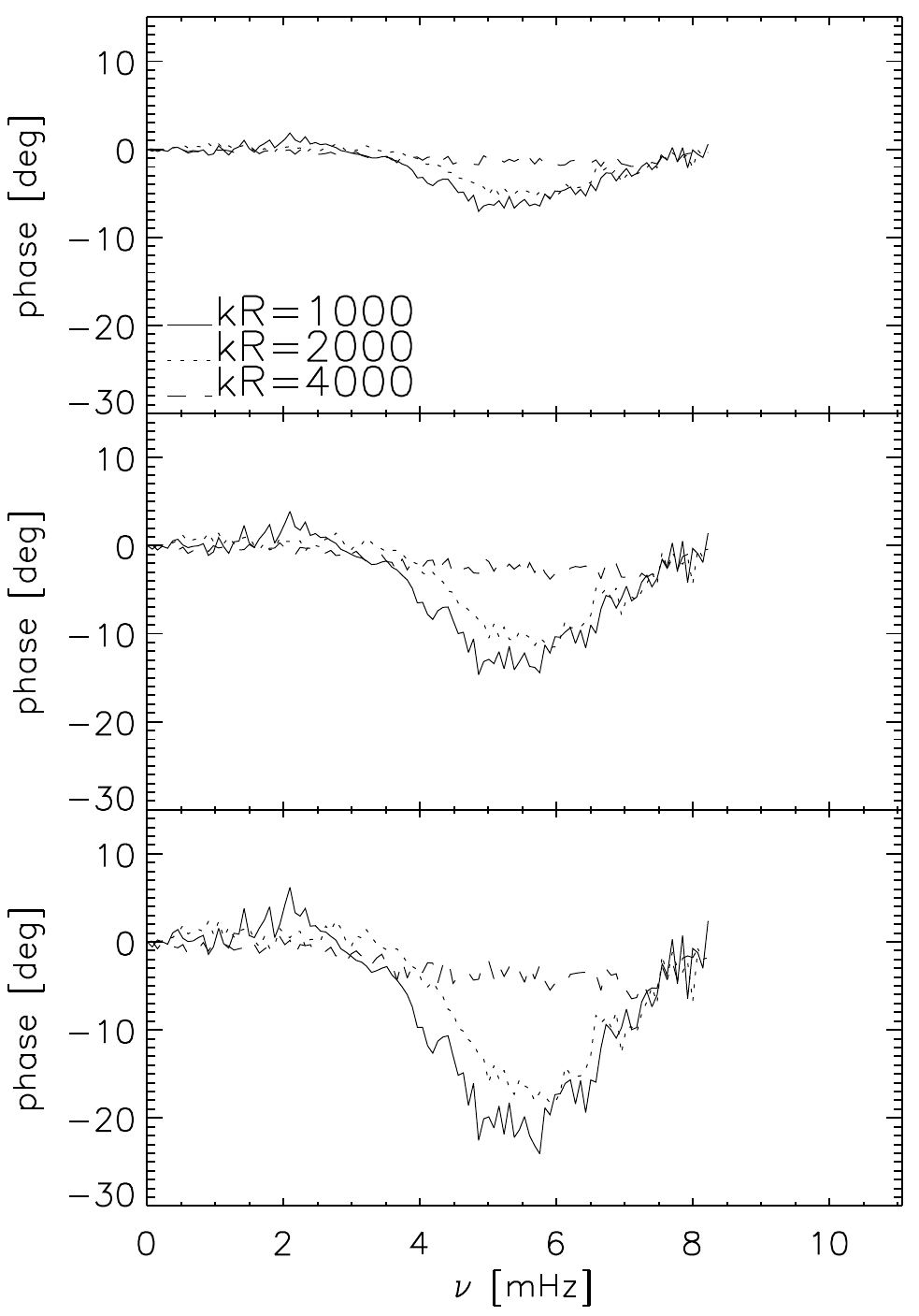} }
\caption{Phase difference of the vertical velocity [$v_z$] between 
the layer at 92 km above the surface and several other layers in 
\textsf{STAGGER} simulation
atmosphere:
(a) $\phi (118 \,\mathrm{km}) - \phi (92 \,\mathrm{km})$,
(b) $\phi(144 \,\mathrm{km}) - \phi(92 \,\mathrm{km})$, and 
(c) $\phi(170 \,\mathrm{km}) - \phi(92 \,\mathrm{km})$. 
The slices at $k \mathrm{R}_\odot=1000$, $2000$, and $4000$ are shown 
in the right panel.
}\label{fig:syn_phasedif_Vz} 
\end{figure}

We choose four layers [92 km, 118 km, 144 km, and 170 km]
and calculate the phase differences of the vertical-velocity field among these layers.
The first three are close to the contribution layers
for the bulk velocities of the average-wing, 
the HMI-algorithm, and the line-center Dopplergrams,
which are estimated by the correlation coefficients between the 
vertical velocity in the atmosphere and the synthetic Doppler velocities
(see Table \ref{tab:ccmaxheight}).
Figure \ref{fig:syn_phasedif_Vz} shows the phase differences between these layers.

\subsubsection{Summary of Phase Difference}
\begin{figure}[tbhp]
\centerline{\includegraphics[width=0.8\textwidth]{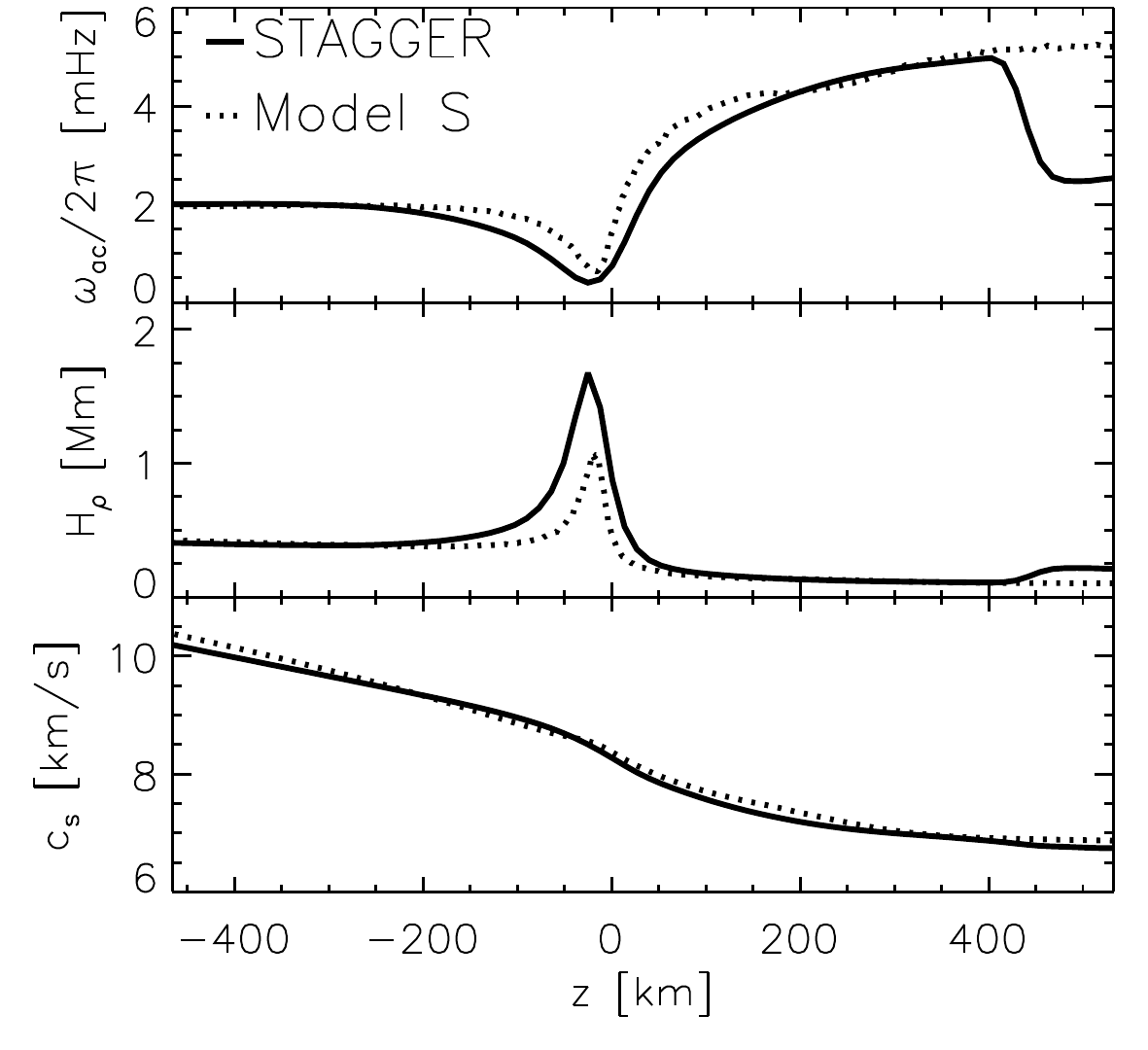}}
\caption{Acoustic cutoff frequency 
[$\omega_{\mbox{ac}} \equiv c_\mathrm{s}/(2H_\rho)$] (top), 
density scale height [$H_\rho$] (middle), and 
sound speed [$c_\mathrm{s}$] (bottom) profiles of \textsf{STAGGER} atmosphere (solid) and
Model S (dotted).}\label{fig:wac}
\end{figure}

Generally, above the photospheric acoustic-cutoff frequency, 
the phase differences found in the HMI observational data 
and the \textsf{STAGGER} data have similar trends, although 
the acoustic-cutoff frequency of \textsf{STAGGER} seems lower than that of the Sun.
Figure \ref{fig:wac} shows the acoustic-cutoff frequency profiles 
near the surface calculated using the mean \textsf{STAGGER} 
atmosphere and the Model S atmosphere.
Here the acoustic-cutoff frequency is defined as 
$\omega_{\mbox{ac}} =c_\mathrm{s}/2H_\rho$, where $H_\rho$ is the density scale height.
The profiles of $c_\mathrm{s}$ and $H_\rho$ are also plotted in Figure \ref{fig:wac}.
The height $z=0$ is defined as the layer where $\tau_{5000 \mbox{\AA}}=1$ 
for each profile (see Figure \ref{fig:prof_atmo}).
The acoustic-cutoff frequency is slightly lower in 
the \textsf{STAGGER} atmosphere, which is consistent with 
what was indicated by the phase difference distributions
(Figures \ref{fig:obs_phasedif} and \ref{fig:syn_phasedif}). 
However, there is a dip of the acoustic cutoff in the top layers 
in \textsf{STAGGER} atmosphere. 
This is due to adding some Newtonian cooling to the energy equation 
near the top boundary to improve computational stability.

Since from the observations we usually cannot see the layer with a constant 
geometrical height but the layer with a constant optical depth, we expect that
the phase differences of the vertical velocity between the 
two iso-optical depth surfaces (Figure \ref{fig:syn_phasedif_Vz_isotau}) 
are more like the observational phase differences 
(Figure \ref{fig:obs_phasedif})
or the phase differences between two synthetic Dopplergrams 
(Figure \ref{fig:syn_phasedif}),
rather than the phase differences of the vertical velocity 
between two iso-geometrical height layers (Figure \ref{fig:syn_phasedif_Vz}).
However, they all look similar to each other, except for Figure \ref{fig:syn_phasedif}b. 
All of the \textsf{STAGGER} phase difference maps have some signatures 
that are not found in the observation, however.
The signatures are located around the boundary of the 
acoustic- and gravity-wave regimes, and in Figure \ref{fig:syn_phasedif}b 
the phase difference is negative and about -20$^{\circ}$, while in other 
figures the phase difference is positive and less than 5$^{\circ}$.
What makes these positive- or negative-value ridges there is still unknown.

In the gravity-wave regime, namely in the lower frequency range, 
all the phase differences of the \textsf{STAGGER} simulation datasets 
show little atmospheric gravity-wave signals,
which are seen clearly in the observation data as was mentioned in Section \ref{sec:phase_obs}.
In the power maps (Figure \ref{fig:obs_syn_power}),
some signatures are seen in the gravity-wave regime
both in the observational Doppler-velocity datasets and synthetic 
(\textsf{STAGGER}) Doppler velocity datasets, and this 
phase-difference signature of atmospheric gravity waves was reported,
\textit{e.g.}, by \inlinecite{2008ApJ...681L.125S} for \textsf{CO$^5$BOLD} 
simulation datasets and by \inlinecite{2009ASPC..415...95S} in observations.
One of the differences between \textsf{STAGGER} and \textsf{CO$^5$BOLD} 
simulations is the extent of the atmosphere. The upper boundary lies 
around 550 km above the surface in \textsf{STAGGER} atmosphere and 
around 900 km in the \textsf{CO$^5$BOLD}atmosphere.
Since there is convection in both simulations, 
there is power in the lower frequency range, \textit{i.e.}, in the gravity wave regime,
but for atmospheric gravity waves the atmospheric extent (about 550 km)
of \textsf{STAGGER} might not be sufficient. 
Another possible interpretation is that
radiative damping of the short-wavelength waves in the \textsf{STAGGER} simulation 
is stronger than in the Sun. 

Note that the resolution of the \textsf{STAGGER} data  used here is the full
resolution, but we checked that the PSF does not change the general trend 
of these phase differences.

\section{Conclusion and Discussion}\label{sec:Conclusions}

We confirm that we can obtain multi-height line-of-sight velocity information 
in the solar atmosphere from SDO/HMI filtergrams.
We suggest average-wing and line-center Dopplergrams
as well as the general pipeline product, (first) HMI-algorithm Dopplergrams
as rather robust multi-height velocity datasets
among the several Doppler velocities defined  in Section \ref{sec:makeDop}
based on the estimate of the contribution layer heights and the
availability of the observables.

In general, multi-height Doppler observations have the potential 
to help constrain the height variations of 
the $p$-mode eigenfunctions in the solar atmosphere 
(see \inlinecite{2012ApJ...760L...1B}, for applications).  
In addition, multi-height Doppler observations may be helpful 
for distinguishing convective motions from oscillations, which in turn may be 
useful to improve the signal-to-noise ratio in helioseismology studies.

We estimate the contribution layer heights of several synthetic 
Doppler velocities 
computed using numerical convection simulation datasets, 
\textsf{STAGGER} and \textsf{MURaM}. 
Although the contribution layer is rather broad, 
the contribution layer heights of 
the average-wing and the line-center Dopplergrams
are 30\,--\,40 km lower and 30\,--\,40 km higher compared to 
the standard HMI-algorithm Dopplergrams, respectively. 
Note that the height difference between these Dopplergrams is
70 km at most, which is about half, at best, of the scale height 
around the surface (see Figure \ref{fig:wac}).
We can obtain multi-height information from these observables, but
since we use the filtergrams taken around a single absorption line, 
the height difference is relatively limited.

HMI observations show clear phase differences between 
these Dopplergrams at frequencies above the acoustic cutoff frequency.
The height difference estimated by the
response functions is consistent with 
the one estimated by the phase differences.
HMI observation data also show a clear signature of 
atmospheric gravity waves in the lower-frequency ranges,
while \textsf{STAGGER} simulation data have only a weak signature. 

Although in this study we limited ourselves to quiet-Sun data 
for the sake of simplicity, multi-height velocity information 
in magnetic regions is also of great interest. 
Since the spectral-line shape is changed not only by the velocity fields 
but also by magnetic field, it is not straightforward to 
analyze such observations and one would need 
radiative-transfer calculations including the effect of 
magnetic field.

The formation layer of the SDO/AIA 1600 \AA \ and 1700 \AA \ passbands 
are estimated in the lower chromosphere around 430 and 360 km above the surface 
(see \opencite{2005ApJ...625..556F}).
Multi-height Doppler observations from SDO/HMI 
either alone or together with SDO/AIA or 
\textit{Interface Region Imaging Spectrograph} (IRIS) observations will 
potentially be also useful to understand how much wave energy is 
transported in the atmosphere and corona
(\textit{e.g.} the review by \opencite{2010LRSP....7....5R}).

%
\begin{acks}[Acknowledgments]
BL acknowledges support from IMPRS Solar System School.
BL computed the line profiles from the \textsf{STAGGER} data cubes 
and the response functions using the \textsf{SPINOR} code.
We thank Michiel van Noort, Thomas Straus, and Jesper Schou for helpful
discussions and comments. 
KN and LG acknowledge support from EU FP7 Collaborative Project 
``Exploitation of Space Data for Innovative Helio-and Asteroseismology"
(SPACEINN). 
LG acknowledges support from 
DFG SFB 963 ``Astrophysical Flow Instabilities and Turbulence" (Project A1). 
The HMI data used are courtesy of NASA/SDO and the HMI science team.  
This work was carried out using the data from the SDO HMI/AIA Joint Science
Operations Center Data Record Management System and Storage Unit
Management System (JSOC DRMS/SUMS). 
The NSO/Kitt Peak FTS data used here were produced by NSF/NSO. 
RS acknowledges support by NASA grant NNX12AH49G and 
NSF grant AGS1141921.
The \textsf{STAGGER} calculations were performed on the
Pleiades cluster of the NASA Advanced Supercomputing Division at Ames
Research Center. 
The German Data Center for SDO (GDC-SDO), funded by the
German Aerospace Center (DLR), provided the IT infrastructure to process 
the data.
\end{acks}


\appendix

\section{Conversion of the Doppler Signal Made of Pairs of Filtergrams into Doppler Velocities} \label{sec:sim_DopsigFit}

\begin{figure}
\centerline{\hspace*{0.015\textwidth}
\includegraphics[width=0.5\textwidth]{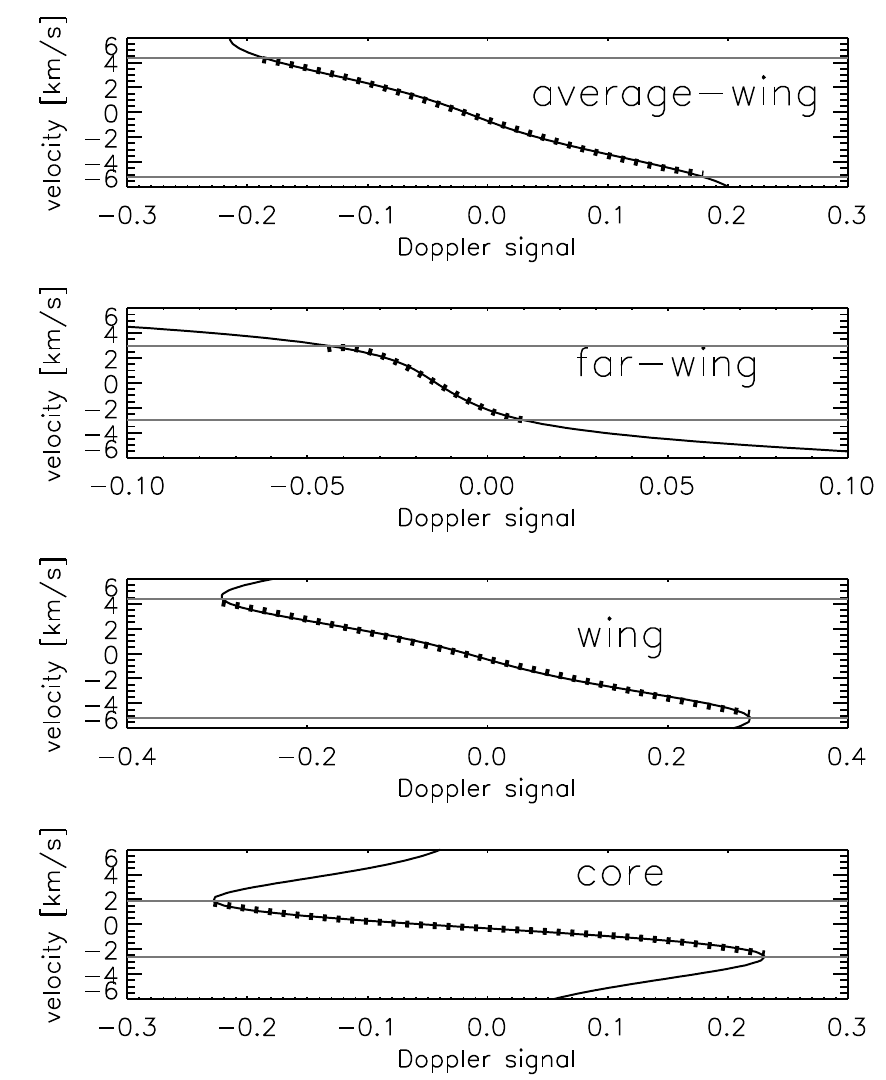}
\hspace*{-0.03\textwidth}\includegraphics[width=0.5\textwidth]{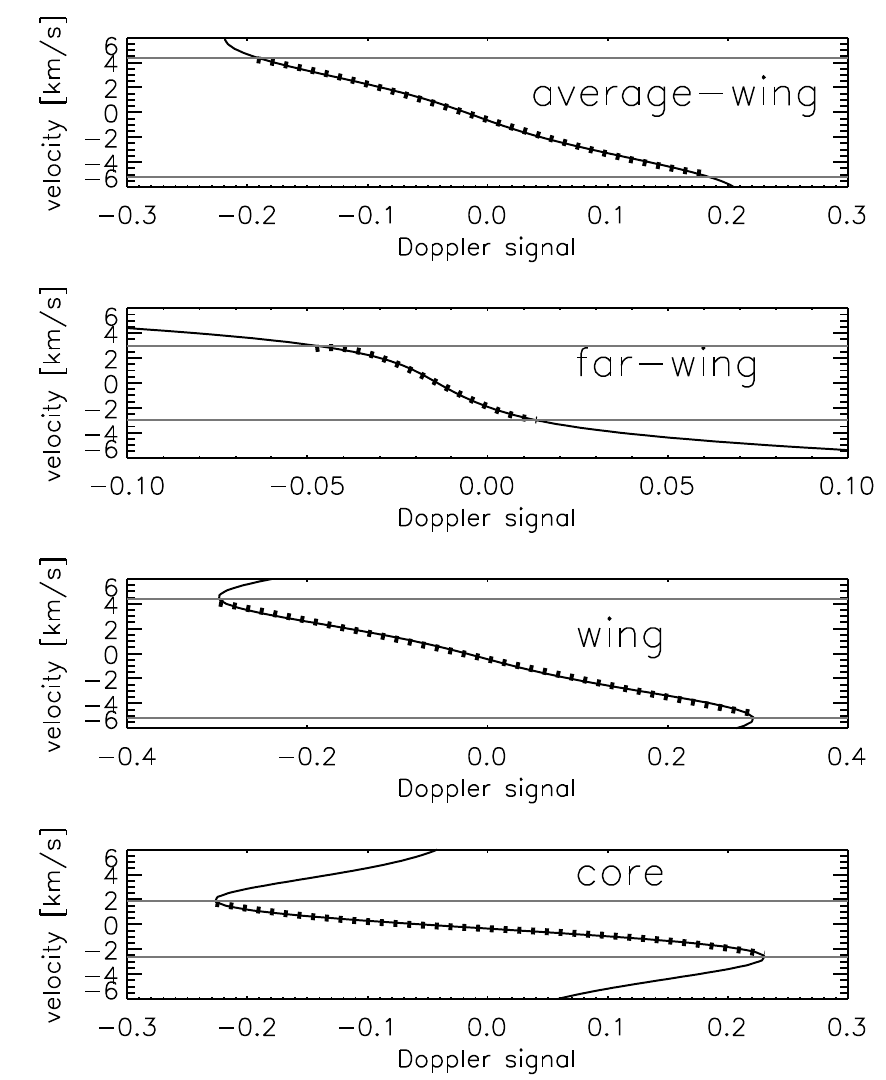}}
\vspace{-0.65\textwidth}
 \centerline{\Large \bf     
      \hspace{0.05\textwidth}   \color{black}{(a)}
      \hspace{0.4\textwidth}  \color{black}{(b)}
         \hfill}
\vspace{0.65\textwidth}
\centerline{\includegraphics[width=0.5\textwidth]{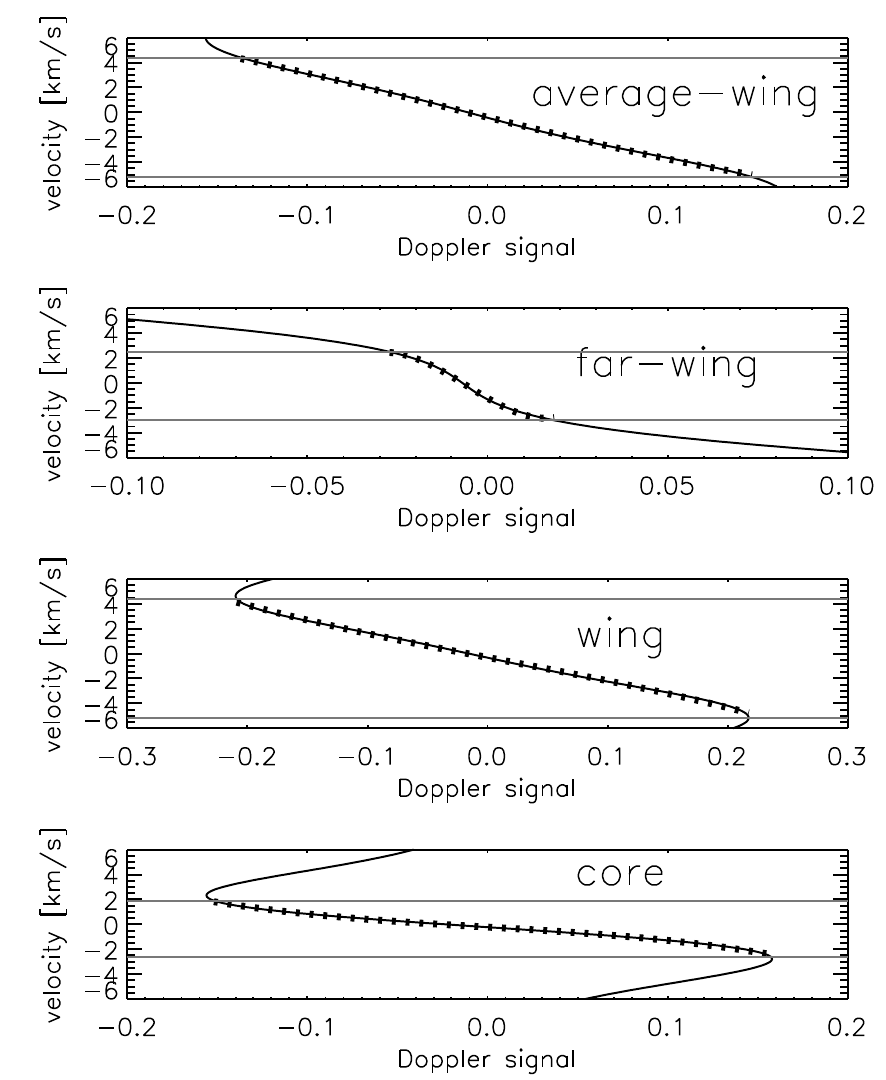}}
\vspace{-0.65\textwidth}
 \centerline{\Large \bf     
      \hspace{0.25\textwidth}   \color{black}{(c)}
         \hfill}
\vspace{0.65\textwidth}
\caption{Doppler-signal changes 
calculated using (a) the \textsf{STAGGER} average line profile, 
(b) the \textsf{MURaM} average line profile, and (c) the HMI reference line profile.
The data points between the solid horizontal lines are used for the fitting, and
the fitted third-order polynomial functions are plotted as dots.
Positive velocity indicates upward velocity, which causes a blue-shift of the line.}
\label{fig:sim_signal_fits}
\end{figure}

For the calculation of the core, wing, far-wing, and average-wing Doppler velocities
(Doppler velocity 1 in the list in Section \ref{sec:DopDef}), 
we convert the Doppler signal [$D_\mathrm{br}$] into the line-of-sight velocity, 
using the parameters we obtain in the following manner:
i)~calculate the Doppler shift of the line core from $-10$ km s$^{-1}$ to $+10$ km s$^{-1}$,
ii)~shift the whole average line profile by the amount of the line-core Doppler shift, 
and calculate the filtergrams by convolving the shifted line profile with the filter profiles,
iii)~calculate the Doppler signals [$D_\mathrm{br}$] for each velocity shift,  and
iv)~fit the velocity to a third-order polynomial function of $D_\mathrm{br}$ 
in a certain range of the velocity.
The fitting range is chosen so that the third-order polynomial fitting seems 
reasonably good. 
Figure \ref{fig:sim_signal_fits} shows how the Doppler signals change as the 
velocity shift changes. The fitting curves are plotted as well.
Figure \ref{fig:sim_signal_fits} clearly shows the saturation of the Doppler signal
with the significantly large Doppler shift. This limits the range useful for the fitting. 

\begin{figure}
\centerline{\includegraphics[]{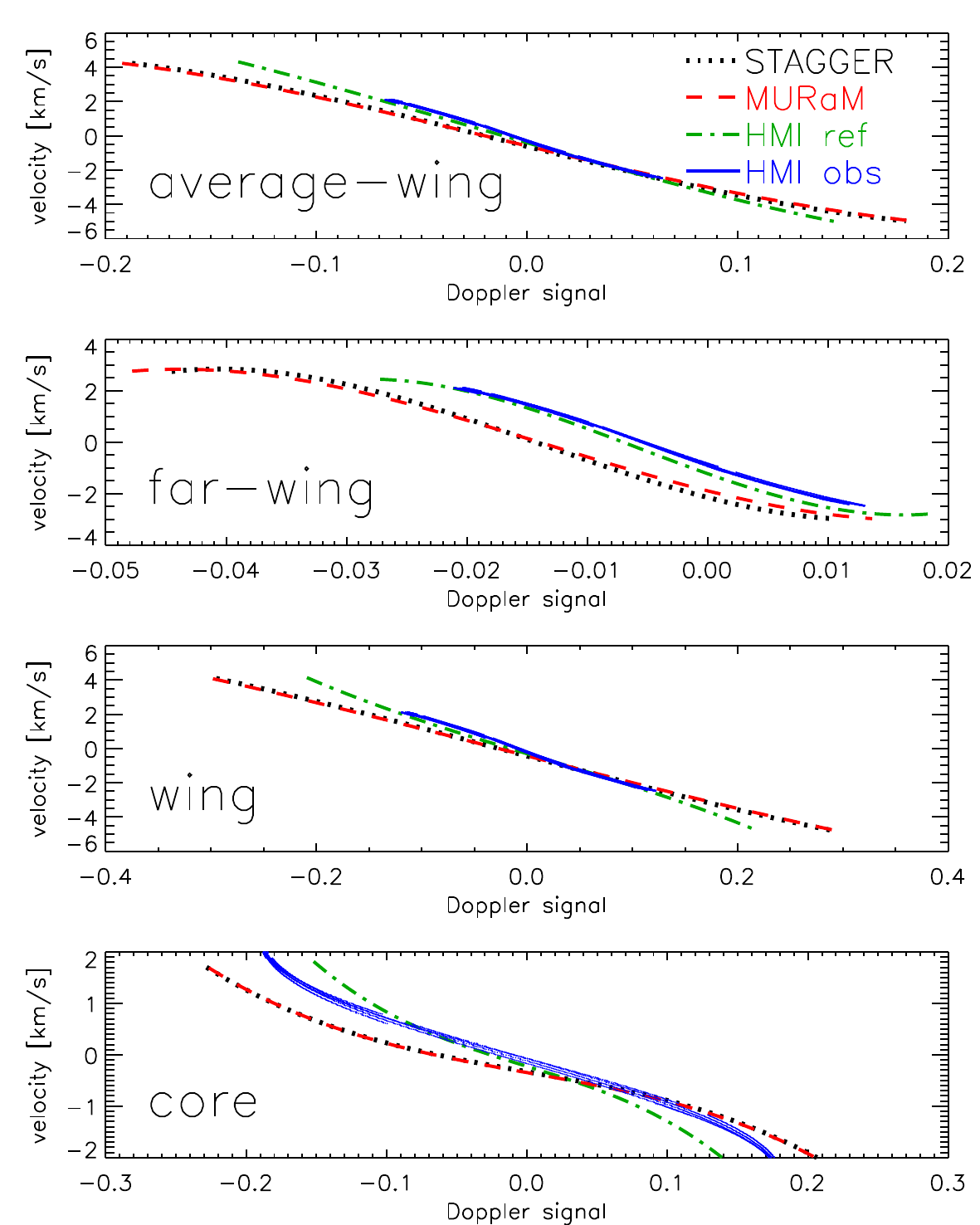}}
\caption{The third-order polynomial fits to the Doppler signals obtained by the fitting 
to convert the signals to the velocities (black dotted, red dashed, 
and green dash-dotted curves for
each line profile) as shown in Figure \ref{fig:sim_signal_fits}. 
Blue dots are the SDO line-of-sight velocity 
plotted against the mean Doppler signals over 
the average of the 30-degree-square field of view near the disk center.
Positive velocity indicates upward velocity.
}\label{fig:sim_Dopsig_polyfit} 
\end{figure}

The conversion parameters depend on the line profile.
The line profiles obtained from \textsf{STAGGER} and \textsf{MURaM} datasets 
and the HMI reference line profiles are different (see Figure \ref{fig:prof_line}), 
and we calculate the fitting parameters for each line profile.
For comparison, the polynomials obtained by the fitting are overplotted in
Figure \ref{fig:sim_Dopsig_polyfit}.
This figure also includes scatter plots of HMI observation Doppler signals averaged over 
a certain field of view (see below) and
the SDO line-of-sight motion towards the Sun. The observation data 
were obtained on 22--24 January 2011. The 30-degree-square quiet region 
near the disk center is tracked at the Carrington rate by 
using \textsf{mtrack} (\opencite{2011JPhCS.271a2008B}).

Note that \inlinecite{2013ASPC..479..429N} used the spacecraft 
motion to compute these conversion curves.
The spacecraft motion is limited to $\approx 3.5 \ \mathrm{km \ s}^{-1}$ 
(\opencite{2012SoPh..275..229S}) and thus
the velocity range of the fitting curve available by this method is limited. 
Therefore, instead, here we use the average line profiles 
(for \textsf{STAGGER} and \textsf{MURaM} datasets) or HMI reference profiles (for observation datasets)
to obtain larger velocity ranges.


%
%
\bibliographystyle{spr-mp-sola}
\bibliography{sdo,localhelios,helioseismology,code,excitation}  
%
%
%
%

\end{article} 
\end{document}